\documentstyle[11pt,aaspp4]{article}

\newcommand{\msun}{\,\hbox{M$_{\odot}$}}
\newcommand{\pcmcu}{\,\hbox{cm}$^{-3}$}
\newcommand{\pcmsq}{\,\hbox{cm}$^{-2}$}
\newcommand{\av}{\hbox{A$_{\rm V}$}}
\newcommand{\lsun}{\,\hbox{L$_{\odot}$}}
\newcommand{\wsqm}{\,\hbox{\hbox{W}\,\hbox{m}$^{-2}$}}
\newcommand{\wsqma}{\,\hbox{\hbox{W}\,\hbox{m}$^{-2}$\,\hbox{arcsec}$^{-2}$}}

\input psfig

\begin{document}

\title{Shock Excitation in Interacting Galaxies: Mkn\,266}
\author{R. Davies}
\affil{Max-Planck-Institut f\"ur extraterrestrische Physik, Postfach
1603, 85470 Garching, Germany}
\authoremail{davies@mpe.mpg.de}

\author{M. Ward}
\affil{Department of Physics \& Astronomy, University of Leicester,
University Road, Leicester, \newline LE1 7RH, UK}

\and 

\author{H. Sugai}
\affil{Department of Astronomy, Kyoto University, Sakyo-ku, Kyoto
606-8502, Japan}

\begin{abstract}
We present near infrared data on the luminous interacting system
Mkn\,266 (NGC\,5256), comprising 2\,$\mu$m continuum, and Br$\gamma$ and
1-0\,S(1) emission line images, together with K-band spectra.
We have fitted stellar templates to the continuum, allowing us to
account for all the stellar features and hence detect even faint
gas excitation emission lines, including 8 and 11 H$_2$ lines in the SW
and NE nuclei respectively.
Population diagrams for the excited H$_2$ molecules indicate that most
of the 1-0\,S(1) in each of the nuclei has a thermal origin.
We discuss this with reference to the observed morphologies, especially
that of the 1-0\,S(1) line.
In particular, the core of 1-0\,S(1) in the NE nucleus is more compact
than the 2\,$\mu$m continuum; while in the SW nucleus the 1-0\,S(1) 
is significantly offset by 500\,pc from the continuum (and other)
emission.
Lastly we address the issue of the region midway between the two
nuclei, where previously a strong source of radio continuum has been
observed.
These results are set in the context of interacting galaxies where
shock excited emission might be expected to occur as a direct
consequence of the interaction.
\end{abstract}


\clearpage

\section{Introduction}
\label{sec:intro}

Prompted by the discovery from CO observations that the bulk of the
molecular gas in close mergers can quickly congregate between the
nuclei (Gao et al. 1997, Scoville et al. 1997)\nocite{gao97,sco97},
we have begun to look at number of such systems, imaging the 1-0\,S(1)
and  Br$\gamma$ lines since the morphologies of these with respect to
the continuum and each other can provide important information.
The rationale is that while star formation models allow a quantitative
analysis of the 
observed fluxes, the results can be quite misleading if the
distribution of the emitting regions is not considered.
Compare, for example, an object in which the Br$\gamma$ and 2\,$\mu$m
continuum come from the same region (in which a single burst of star
formation is possible), to one in which they are offset from each
other, {\it requiring} two distinct epochs of activity.
A classic example is given by \nocite{gon97}
Gonz\'alez Delgado et al. (1997)\nocite{gon97} 
who found both Wolf-Rayet and Ca{\sc ii}
features in the super star cluster A of NGC\,1569, implying the
simultaneous presence of hot massive stars and red supergiants.
Their intepretation required the existence of non-coeval stellar
populations in the same stellar cluster.
The difficulty was resolved when De~Marchi et al. (1997)\nocite{mar97}
found from HST imaging that this was actually 2 distinct clusters.
Such high resolution is not always required, and
in a small sample of blue compact dwarf galaxies observed by Davies et
al. (1998)\nocite{dav98b}, the Br$\gamma$, 1-0\,S(1), and 2\,$\mu$m continuum all had
strikingly different morphologies on scales of a few arcsec
($\sim$100\,pc).
In such cases a simple interpretation of the emission in
terms of just star formation is not valid.

Another issue which has posed a significant barrier to extragalactic
work is the detection of the fainter H$_2$ lines, which is essential
to determine the fractions of H$_2$ which are thermally and
non-thermally excited.
Previously, except in a few cases such as NGC\,6240 (Sugai et al. 1997b)\nocite{sug97b}
which has extremely bright 1-0\,S(1), this has typically relied on very
few line 
ratios, often given only as upper limits and extracted from rather
large apertures (eg Goldader et al. 1997)\nocite{gol97b}.
It is crucial to detect lines from the v=2 (or higher)
bands: a single thermal model can nearly always provide a reasonable
fit to lines from the v=1-0 transitions.
It is also important to have more than one v=2-1 line,
since ratios involving only a single such transition can be ambiguous:
for temperatures in the range 1500--2500\,K, the thermal
2-1\,S(1)/1-0\,S(1) ratio varies from 0.03 to 0.15.
Thus if a ratio (or upper limit) of 0.15 is observed,
while the 1-0\,S(1) is constrained to be $>75$\% thermal,
the {\it total} H$_2$ cooling can only be attributed to somewhere in
the range 35--100\% thermal depending solely on a fairly small change 
in thermal excitation temperature.
With the increasing sensitivity of near-infrared detectors it is now
possible to use small apertures and obtain the high
signal-to-noise spectra required to detect higher vibrational lines.
Additionally, fitting the continuum with stellar templates, rather than
using a power-law or polynomial fit, is
necessary in order to take account of the stellar features and allow
weak lines from the excited gas, such as H$_2$ lines, to be measured
reliably.

We address these issues in this paper, in which we present a detailed
analysis of near infrared emission line images and spectra.
We focus on Mkn\,266, a luminous infrared galaxy 
($L_{\rm IR} = 3\times10^{11}\,L_\odot$) 
at a distance of 115\,Mpc (1\arcsec\ = 560\,pc).
It has 2 prominent nuclei 10\arcsec\ apart, with a common envelope, 
suggestive of a merger 
(Mazarella et al. 1988,\nocite{maz88} hereafter MGAH).
There have been very few detailed studies of Mkn\,266, the two most
decisive being those of MGAH and Wang et al (1997)\nocite{wan97}.
The former authors presented an H$\alpha$ map which revealed 
arcs that may be tidal tails or remnant spiral structure.
They also mapped the radio continuum, which
appears similar to the 2\,$\mu$m continuum but with a strong region of
extended emission halfway between the nuclei.
The latter study found evidence for a superwind in their X-ray imaging and
spectroscopic data.

\section{Observations and Data Reduction}
\label{sec:obs}

\subsection{Fabry-Perot Images}
\label{sec:obs:fp}

Images of the 2\,$\mu$m continuum, and 1-0\,S(1) and Br$\gamma$ emission lines were
obtained with the UKIRT 3.8-m telescope on Mauna
Kea, on the nights of 12 and 13 May 1998.
A K-band Fabry-Perot etalon (FP), set in a collimated beam, was
used in conjunction with IRCAM3 (a $256\times256$ InSb array) at the
Cassegrain focus.
The scale of 0.286\arcsec\ per pixel gave an unvignetted field
of view of $\sim$60\arcsec.
Order sorting was achieved with cooled narrow band filters:
2.4\% FWHM for the Br$\gamma$ line, and 4.25\% FWHM for the 1-0\,S(1) line.
The latter filter is normally used for continuum imaging, but was
required due to the large recession velocity ($>8000$\,km\,s$^{-1}$) of the
galaxy which shifted the line out of the standard filter pass-bands.
The 325\,km\,s$^{-1}$ spectral resolution of the FP and the wavelength
calibration were checked several times during the night using the
2.1171\,$\mu$m line from a Krypton lamp, permitting adjustments
for temperature changes (approx. 40\,km\,s$^{-1}$/\,$^\circ$C).

Due to the relatively small size of Mkn\,266 we imaged it at 2
positions on the detector, symmetrically off-axis from the centre of
the FP (14\arcsec\ North-West and 14\arcsec\ South-East).
The transmitted wavelength varies off-axis as $\delta\lambda
\propto r^2$, reaching $\delta\lambda \equiv 160$\,km\,s$^{-1}$ at $r=30$\arcsec,
hence the difference is only $\delta\lambda = 40$\,km\,s$^{-1}$ at $r=14$\arcsec.
This was taken into account when setting the FP wavelength.
Other concerns include the small 15\,km\,s$^{-1}$ offset for the velocity of the
earth relative to the sun in the direction of the object on that date,
which was not included, and the uncertainty in the heliocentric line
velocity.
Our adopted value was 8385\,km\,s$^{-1}$, the average for the two nuclei
determined from optical spectra by Veilleux et
al. (1995)\nocite{vei95}, resulting in additional offsets of
$\pm25$\,km\,s$^{-1}$ for each nucleus.
Although each of these adjustments is small, they must all be considered:
with a FWHM resolution for the FP of 325\,km\,s$^{-1}$, a total offset of 100\,km\,s$^{-1}$ 
will mean that the line flux is underestimated by 35\%, but a reduction
to only 50\,km\,s$^{-1}$ (similar to the maximum we expect)
results in a net effect on the line flux of less than 10\%.
A similar consideration applies to the line dispersion, which can reduce
the line flux detected if it is much greater than 100\,km\,s$^{-1}$.
Typical corrections are 20\% for 200\,km\,s$^{-1}$, and 35\% for 300\,km\,s$^{-1}$, but it
should be noted that they combine with velocity offsets non-linearly.
Line dispersion effects were not included in the line flux estimation
since the line-widths are not known, although they are expected to be
relatively narrow for starbursts.

Observations were carried out in blocks, each of which had 4
integrations of 3\,minutes per pointing, consisting of: 2 at
the on-line wavelength, and one each for continuum at slightly shorter
and longer wavelengths (shifted with respect to the line centre by -800
and +1000\,km\,s$^{-1}$ respectively).
Standard A-type stars (BS\,4344 and BS\,5142) were observed between each
set of 2 or 3 blocks.
The total on-line integrations were 48\,mins for Br$\gamma$, 
and 120\,mins for 1-0\,S(1).

IRAF was used for all image processing and analysis as described below.
Flatfields were made using opposite positions after dark current subtraction.
These consist of the camera flatfield, and a smooth circularly symmetric
contribution from the FP.
Due to small temperature variations, the latter changed slightly between
successive integrations (at the 1\% level), leaving a residual ripple.
Although this makes little difference to the continuum counts, it is
crucial to the detection of line flux since the {\em peak} line counts
per pixel are only 2\% of the sky background. 
Thus after each frame was dark subtracted and flatfielded, it was also
divided by the median filtered combination of itself rotated every
22.5$^\circ$ about the centre of the ripple pattern.
A constant value for the sky could then be subtracted (rather than
using sky frames which would increase the noise).
The frames were aligned on the NE nucleus since it is more
compact than the SW one, but this could introduce systematic offsets
due to the line emission in the on-line frames.
In order to check that there was no such bias, the
centroid positions between pairs of successive off-line and on-line
images (at the same telescope pointing) were compared.
For the Br$\gamma$ line these were $x=-0.17\pm0.47$\,pixels and
$y=0.05\pm0.27$\,pixels; for 1-0\,S(1) they were $x=-0.10\pm0.12$\,pixels
and $y=0.24\pm0.20$\,pixels ($\rm 1\,pixel=0.286$\arcsec).
For both lines, the mean difference was less than the RMS variation and
systematic effects should not be more than $\sim$0.1\arcsec.

Flux calibration was achieved with the standard stars BS\,4344 (type
A4V, K=6.30) and BS\,5142 (type A3V, K=5.18).
The frames were treated in the same way except that sky frames
were subtracted and there was no need to correct for the ripple pattern.
The profiles of the order sorting filters were included, making as much
as 15\% difference from assuming a simple
boxcar shape due to secondary FP transmission orders in the
wings of the pass-band.
The image resolution was determined from the stars to be
0.77\arcsec\ for the 1-0\,S(1) line and 1.34\arcsec\ for the Br$\gamma$ line.

The continuum image was deconvolved with 30 iterations of the Lucy
algorithm implemented in IRAF, after halving the pixel size.
The iterations were stopped once false structure due to noise peaks
began to show in the fainter emission. 
At this point the stronger emission peaks show only genuine features;
for example, the asymmetry in the NE nucleus in the deconvolved image
is also apparent (just) in the direct image.
The resolution of 0.36\arcsec\ was determined by including at one edge
of the image an extra region with the same noise and a number of
Gaussians convolved with the PSF;
after deconvolution, the recovered and original sizes were compared.

\subsection{Longslit Spectroscopy}
\label{sec:obs:slit}

\begin{figure}
\centerline{\psfig{file=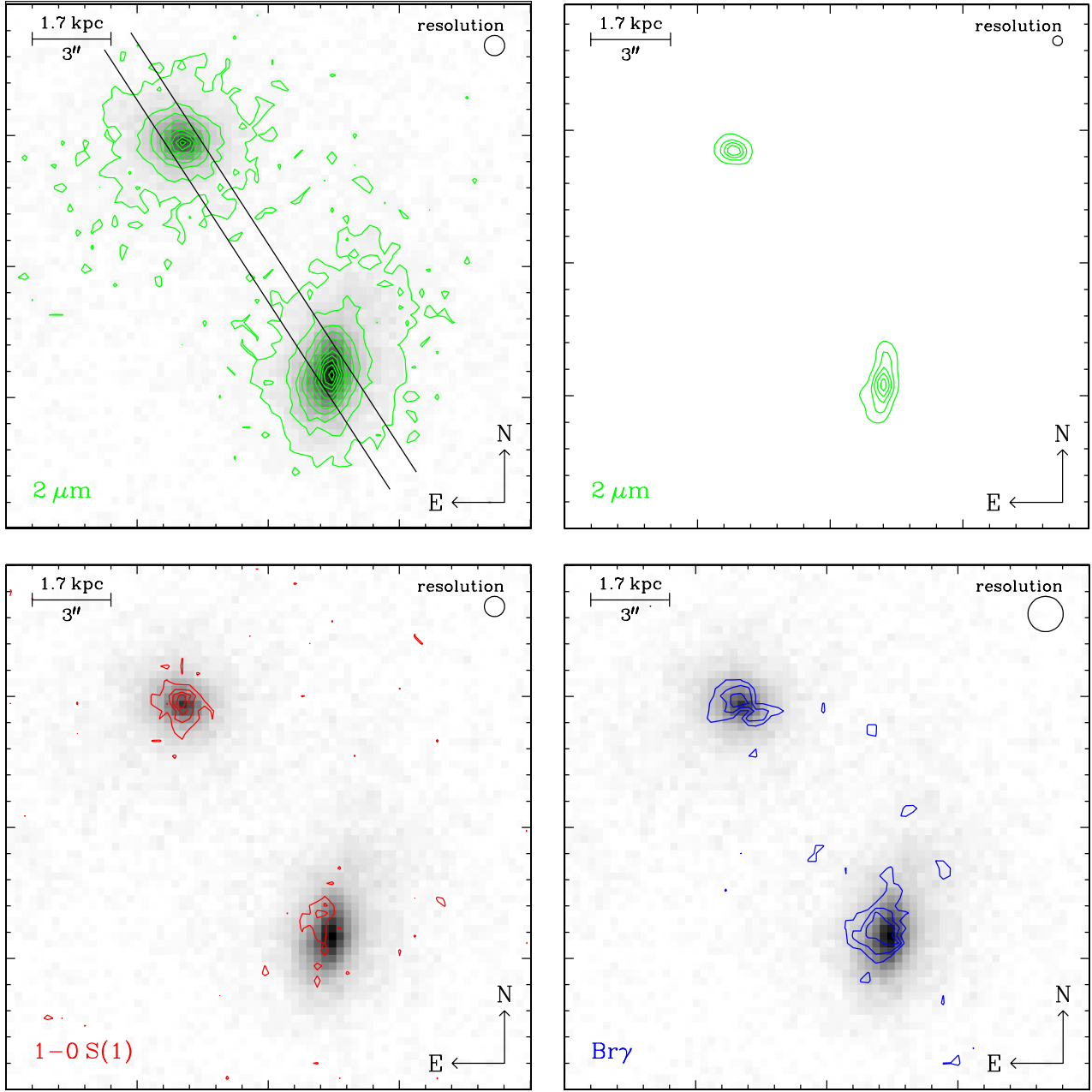,width=18cm}}
\caption{Contour maps of the 20\arcsec\ region around Mkn\,266.
Upper left: 2\,$\mu$m continuum.
Contours are from $2\sigma$\,pixel$^{-1}$ at intervals of 
$6\sigma$\,pixel$^{-1}$, where 
$1\sigma$\,pixel$^{-1} \equiv 0.072$\,mJy\,arcsec$^{-2}$.
Upper right: 2\,$\mu$m continuum after rebinning and deconvolving with 30
iterations of the Lucy algorithm (see Section~1 for details) at
arbitrary linear intervals.
Lower left: 1-0\,S(1) line superimposed on continuum greyscale.
Contours are from $2.5\sigma$\,pixel$^{-1}$ at
intervals of $2.5\sigma$\,pixel$^{-1}$, where 
$1\sigma$\,pixel$^{-1} \equiv 1.6\times10^{-19}$\wsqma.
Lower right: Br$\gamma$ line (smoothed using a $3\times3$ median filter, with
little effect on the resolution) superimposed on continuum greyscale.
Contours are from $2.5\sigma$\,pixel$^{-1}$ at
intervals of $1.5\sigma$\,pixel$^{-1}$, where 
$1\sigma$\,pixel$^{-1} \equiv 2.1\times10^{-19}$\wsqma.
\label{fig:morph}}
\end{figure}

Spectra of Mkn\,266 were obtained as part of the UKIRT service observing
programme on May 1 1999 using the CGS4 spectrometer;
the orientation of the slit, 33$^\circ$ East of North, is indicated on
Fig.~\ref{fig:morph}.
The total integration time was 64\,min.
Frames are sky subtracted, flatfielded, and co-added on-line using CGS4DR;
further data reduction was carried out with IRAF.
The dispersion axis was aligned to pixel rows, and the spatial axis
to pixel columns with reference to an argon arclamp and hence
simultaneously calibrating the wavelength scale.
Standard A stars (BS\,5023 and HD\,105601) were used for flux
calibration and atmospheric correction after interpolating over
Br$\gamma$ and Pa$\alpha$ with Lorentzian profiles and making a small
correction to the same airmass as Mkn\,266.
The spatial resolution was determined from the standard stars to be
1.06\arcsec, well matched to 2-pixel sampling.
Integrated spectra of each nucleus were extracted in order to look at
weak features, and additionally the
spatial extent of the brighter features was considered.
The heliocentric velocities of the two nuclei were measured as
8440\,km\,s$^{-1}$ (NE) and 8390\,km\,s$^{-1}$ (SW), in good agreement with those of
Veilleux et al. (1995)\nocite{vei95}.

\section{Emission Line Images}
\label{sec:images}

\begin{figure}
\centerline{\psfig{file=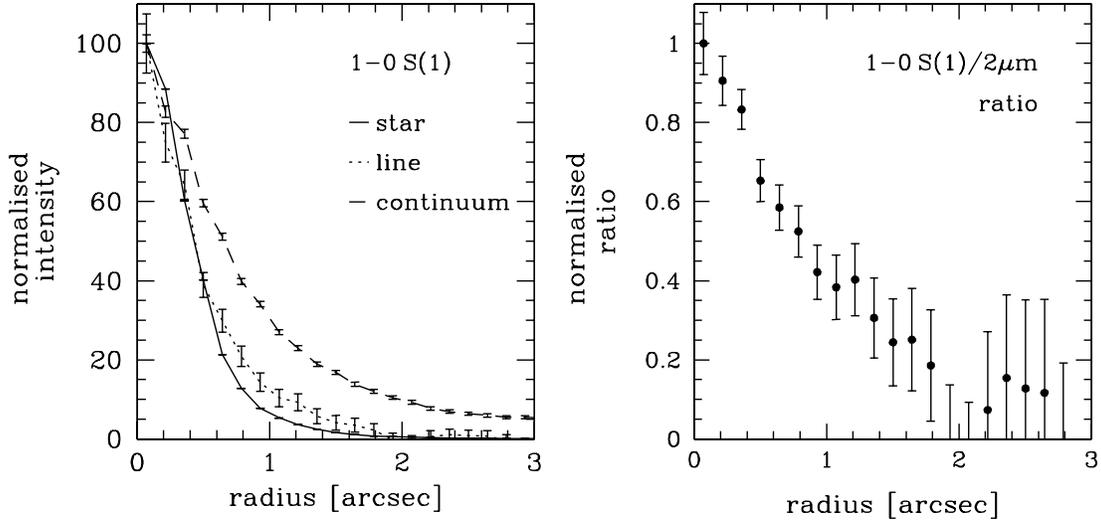,width=16cm}}
\caption{Left: Measured profiles of a star, 2\,$\mu$m 
continuum and 1-0\,S(1) line in the North-East nucleus.
Azimuthally averaged intensity profiles are derived from flux in
concentric annuli.
Right: ratio of 1-0\,S(1)/2\,$\mu$m (ie normalised equivalent width of the
line), decreases as a function of radius.
These diagrams show clearly that the core of the 1-0\,S(1) is
(a) unresolved, whereas the continuum is marginally resolved; and 
(b) has faint wings of emission at 1--2\arcsec.
\label{fig:profile.s1}}
\end{figure}

The continuum image (upper left and, deconvolved, upper right) in
Fig~\ref{fig:morph} shows that the 2 nuclei are separated by
10.5\arcsec, 5.9\,kpc at the distance of the galaxy.
The NE nucleus is resolved with a size of 0.8\arcsec$\times$0.6\arcsec.
As Fig~\ref{fig:profile.s1} shows the 1-0\,S(1) is more compact than
this: comparison via quadrature correction with the continuum would put
its size at $\lesssim 0.45$\arcsec.
This is similar to the extent of the radio source observed at 6\,cm by
MGAH, which was marginally resolved at
a resolution of 0.3\arcsec$\times$0.4\arcsec, and found to have a
spectral index of 0.61$\pm$0.04.

The continuum in the SW nucleus has a morphology very similar to the
radio continuum: central source with extensions to $\sim$1\arcsec\ at
position angles slightly west of North and east of South.
The Br$\gamma$ may also include these regions but the poorer resolution
rules out a definitive statement.
Nevertheless, the presence of a fairly substantive continuum component
argues against the suggestion of MGAH that these are radio
jets from the Seyfert nucleus, and tends towards interpretation in
terms of star formation.
An interesting point is that the high CO index measured in the spectrum by 
Goldader et al. (1997)\nocite{gol97b}, CO$_{\rm ph} = 0.16\pm0.03$,
suggests the presence of late-type supergiants, which would only be
expected in a young stellar population.
Such a conclusion is consistent with our spectra which suggest a 50\%
contribution 
from supergiants in the K-band, twice that of the NE nucleus.
The implied spatial scales on the order of 1\,kpc are consistent with what is
known about circumnuclear star formation occuring at inner Lindblad
resonances around an AGN/starburst.
It should be noted, however, that there is currently no evidence for
the barred potential in a disk which would lead to gas inflow to such
resonances.
Indeed, the H$\alpha$ and optical morphologies (MGAH) are highly perturbed.

The 1-0\,S(1) in this region emission is rather unexpected, the peak
being offset from the continuum by 0.9\arcsec.
As described in Section~\ref{sec:obs}, although the images were
aligned on the NE nucleus, systematic effects cannot account for these
offsets.
The 1-0\,S(1) line map has a higher signal-to-noise and good resolution, so
we can be confident that the offset is real.
For the Br$\gamma$ map a big uncertainty lies in the low signal-to-noise
(peak pixel in the smoothed image is $7.5\sigma$),
and the relatively poor resolution (1.3\arcsec).
We therefore cannot claim any significance in the Br$\gamma$ offset.
Astrometry with the 6\,cm radio map of MGAH suggests that the Br$\gamma$ and
2\,$\mu$m continuum are aligned with the Seyfert nucleus.

\section{Spectra}

\begin{figure}
\centerline{\psfig{file=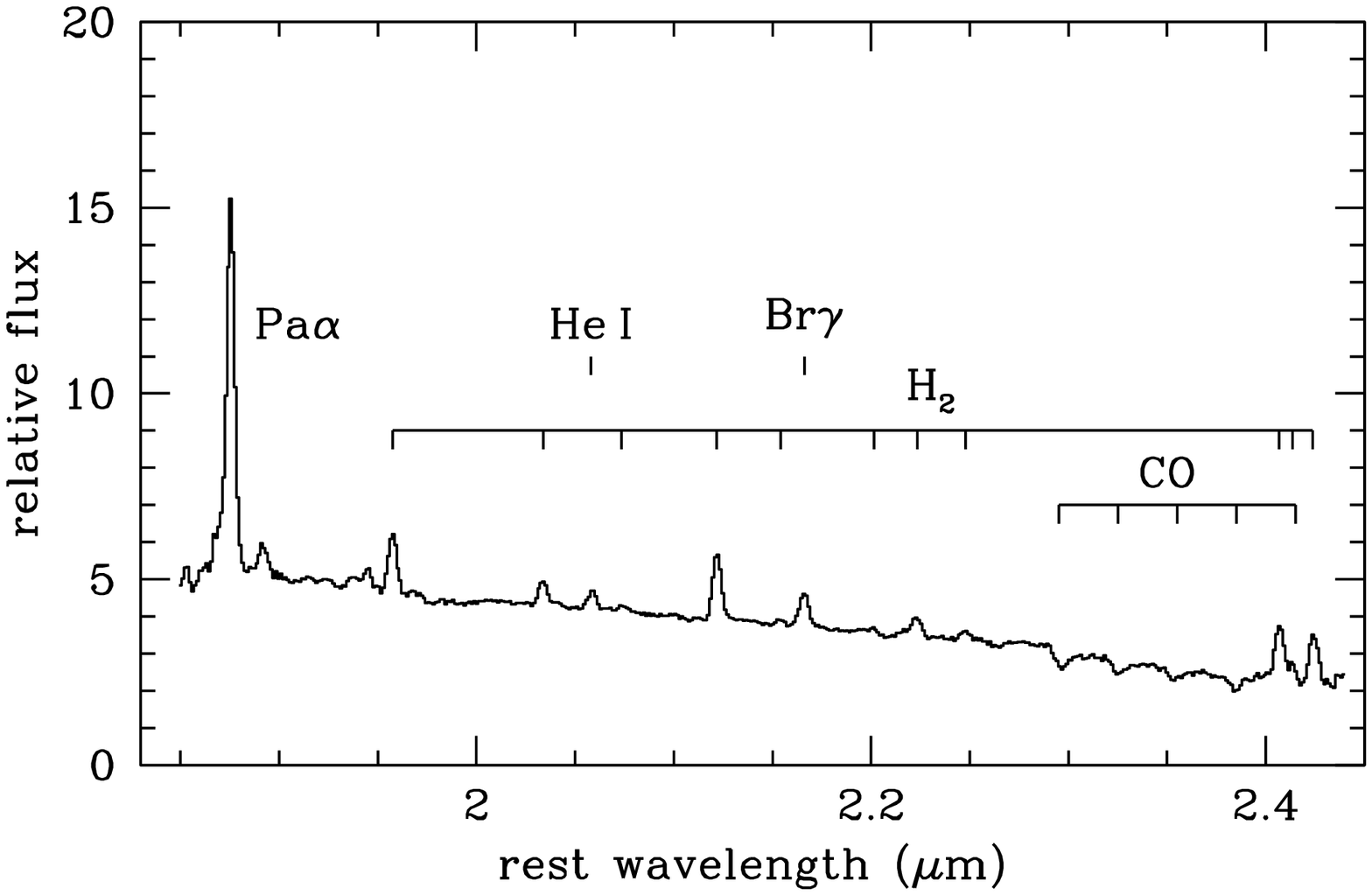,height=9cm,width=15cm}}
\centerline{\psfig{file=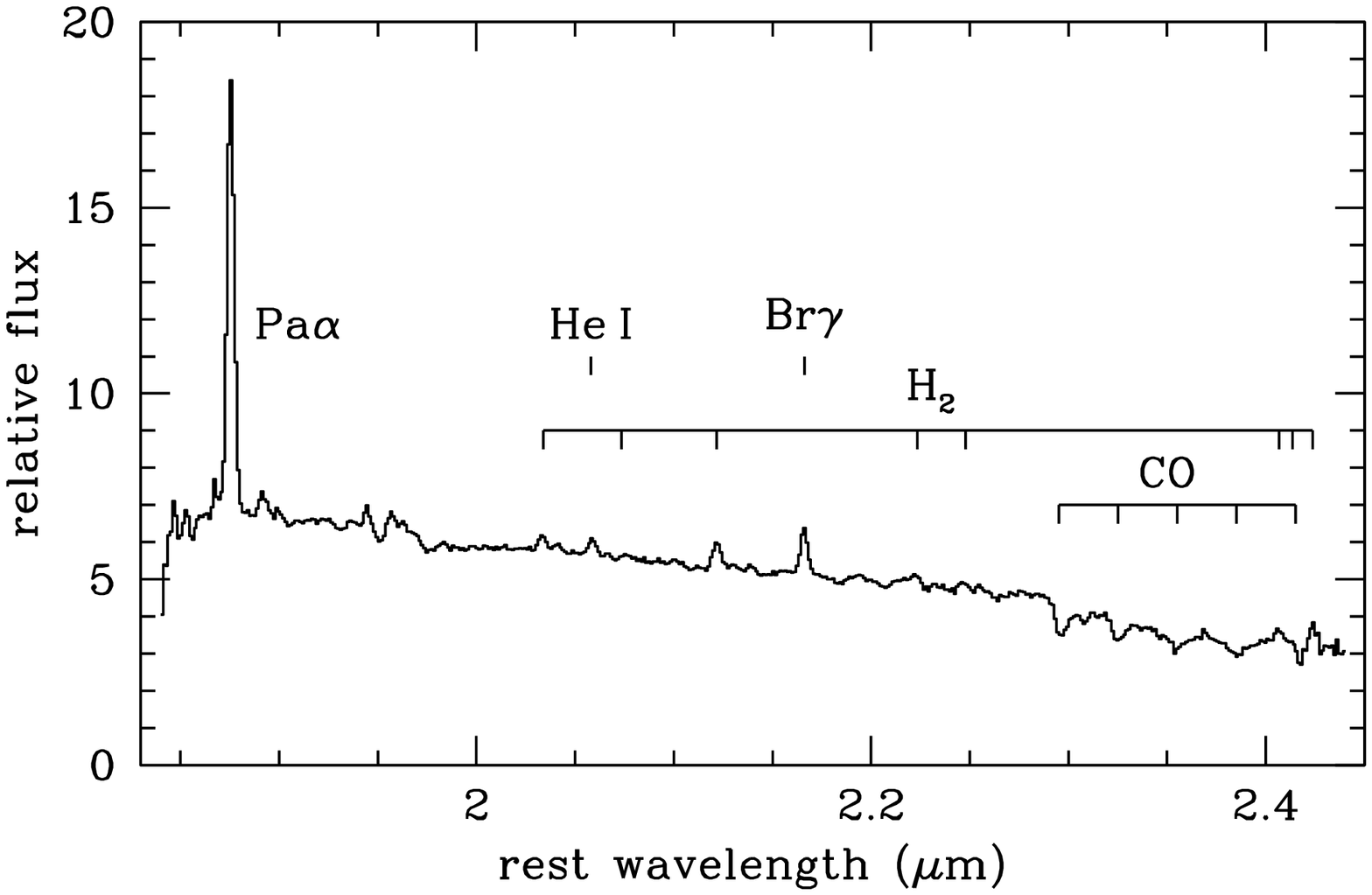,height=9cm,width=15cm}}
\caption{Integrated spectra of the NE (upper) and SW (lower) nuclei of
Mkn\,266.
The important emission and absorption features are indicated.
Fitting and subtraction of the continuum with stellar teplates is
important in order to account for absorption features close to emission
lines (eg the H$_2$ Q-branch).\label{fig:spec}}
\end{figure}

\begin{figure}
\centerline{\psfig{file=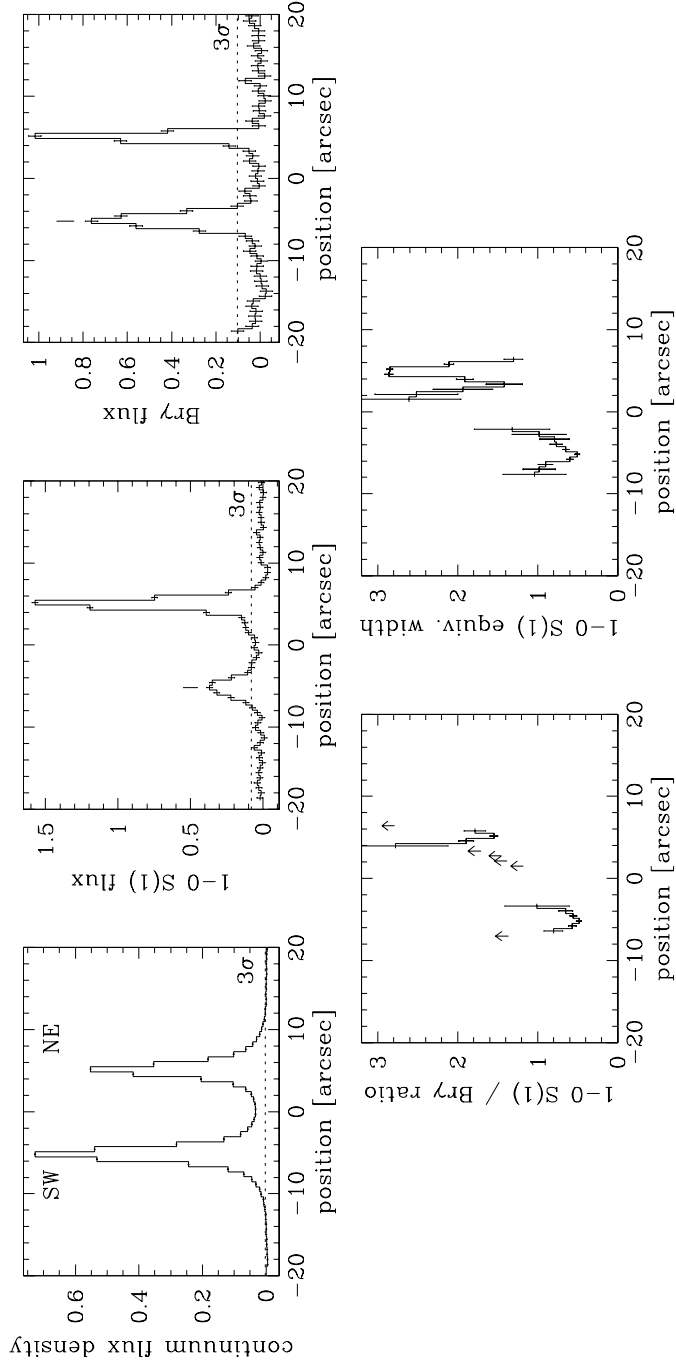,height=20cm}}
\caption{Upper:
2$\mu$m continuum, 1-0\,S(1) flux, and Br$\gamma$ flux,
traced for 40\arcsec\ along the slit across the SW and NE nuclei.
In the latter 2 figures, the vertical bar indicates teh SW continuum
centroid;
unlike the images, no offset is seen between the 1-0\,S(1) and 2$\mu$m
continuum due to the poorer spatial smapling and seeing.
Lower:
ratio of 1-0\,S(1)/Br$\gamma$ and 1-0\,S(1)/2$\mu$m (equivalent width).
Away from the nuclei, both ratios increase.
In the NE nucleus, the initial decrease is the same as that seen in
Figure~2.
\label{fig:profile.s1brg}}
\end{figure}

An integrated spectrum of each nucleus was extracted, covering 5 pixels
(3.05\arcsec) along the slit by the slit width of 1.22\arcsec, and
these are shown in Fig.~\ref{fig:spec}.
Additionally, the spatial extent of some stronger features could be
mapped, as shown in Fig.~\ref{fig:profile.s1brg}.
Clearly the system is far from trivial, and it would be useful to analyse
all the line ratios with high spatial resolution, something which may
be possible with adaptive optics and integral field spectroscopy in the
(near) future (eg see Davies et al. 1999)\nocite{dav99a}.
Here we present an analysis of the detailed integrated spectra together
with some spatial information from both spectra and images.

\subsection{Continuum Fitting \& CO Absorption}
\label{sec:contfit}

In order to recover emission lines which are either faint or
superimposed on absorption features it is important to subtract an
accurate continuum rather than fit a simple line or curve as is often
done.
This has the double benefit of also providing a qualitative estimate of
the stellar population contributing to the 2$\mu$m emission.
We used a minimum chi-square fitting routine with the following
restrictions:
\newline
(1) only regions of the continuum distinct from emission lines were
included in the chi-square estimations;
\newline
(2) a single value for extinction was fitted;
\newline
(3) a hot dust component (eg for an AGN) was not included as it did not
improve the fits.

The CO absorption indicates that late-type giant \& supergiant stars
dominate the continuum at these wavelengths and so we used generic
spectra of these types, 
created by adding several examples of each from the library of
F\"orster-Schreiber (1999)\nocite{for99} and convolving them to the same
resolution as our data.
The individual templates have varying spectral coverages (dependent on
the resolution), resulting in a trade-off between large coverage of
2.00--2.44$\mu$m or a wider variety of stellar types but covering only
2.27-2.40$\mu$m.
The latter range of templates includes M\,I, K\,I, K0\,III, K5\,III,
M0\,III, which cover the range of types required;
adding any other available spectra (eg between K0 and K5 giants) would
not provide any further useful information about the stellar population. 
The former range of types misses out the M0\,III and, more seriously, K\,I.
Nevertheless, both approaches gave the
same qualitative result, and so we 
used the former one in order to provide the longest possible baseline
for subtraction.
The stellar types used were M\,I, K0\,III, and K5\,III, and the best
fits had RMS errors of $\sim$0.06 per data-point (for the scales in
Fig.~\ref{fig:spec}).
The baseline of the fitted continuum was extended down to 1.85$\mu$m by
adding a much lower 
resolution spectrum of a late-type giant star, also reddened with the
same extinction, from the PEGASE database (Fioc \& Rocca-Volmerange
1997)\nocite{fio97}.

The result showed that about 20--25\% of the 2$\mu$m continuum in the
NE nucleus is from late-type supergiants, while the rest is from
giants, and the extinction is A$_{\rm V} \sim 3.2$\,mag;
in the SW nucleus about 50\% is from supergiants, with A$_{\rm V} \sim
4.6$\,mag.

We have additionally tried to constrain the stellar population by
measuring the CO index directly.
We have used two indices, CO$^\prime$ of Goldader et
al. (1997)\nocite{gol97b} in the
2.30--2.34\,$\mu$m interval, and CO$_{\rm sp}$ of Doyon et
al. (1994)\nocite{doy94} in the 2.31--2.40\,$\mu$m range.
Converting these to CO$_{\rm ph}$, using the relations given by
Goldader et al., gives similar values for the two indices.
These values are CO$_{\rm ph} = 0.12$ (NE) and 0.16 (SW).
The latter value verifies that there is a higher supergiant fraction in
the SW nucleus, and is the same as that measured by Goldader et
al. even though they used a much larger 3\arcsec$\times$9\arcsec\
aperture.

\subsection{H$_2$ Excitation}
\label{sec:h2excite}

The process described above removes much of the structure in the continuum,
including such features as Mg (2.14$\mu$m), Na (2.21$\mu$m), Ca
(2.26$\mu$m), etc.
It allows not only weak lines to be measured, but also the Q-branch
H$_2$ lines -- which is useful since
the 1-0\,Q(3) and 1-0\,S(1) lines come from the same
upper level, as do 1-0\,Q(2) and 1-0\,S(0), providing a 
consistency check when plotting a population diagram for the molecules.

\begin{figure}
\centerline{\psfig{file=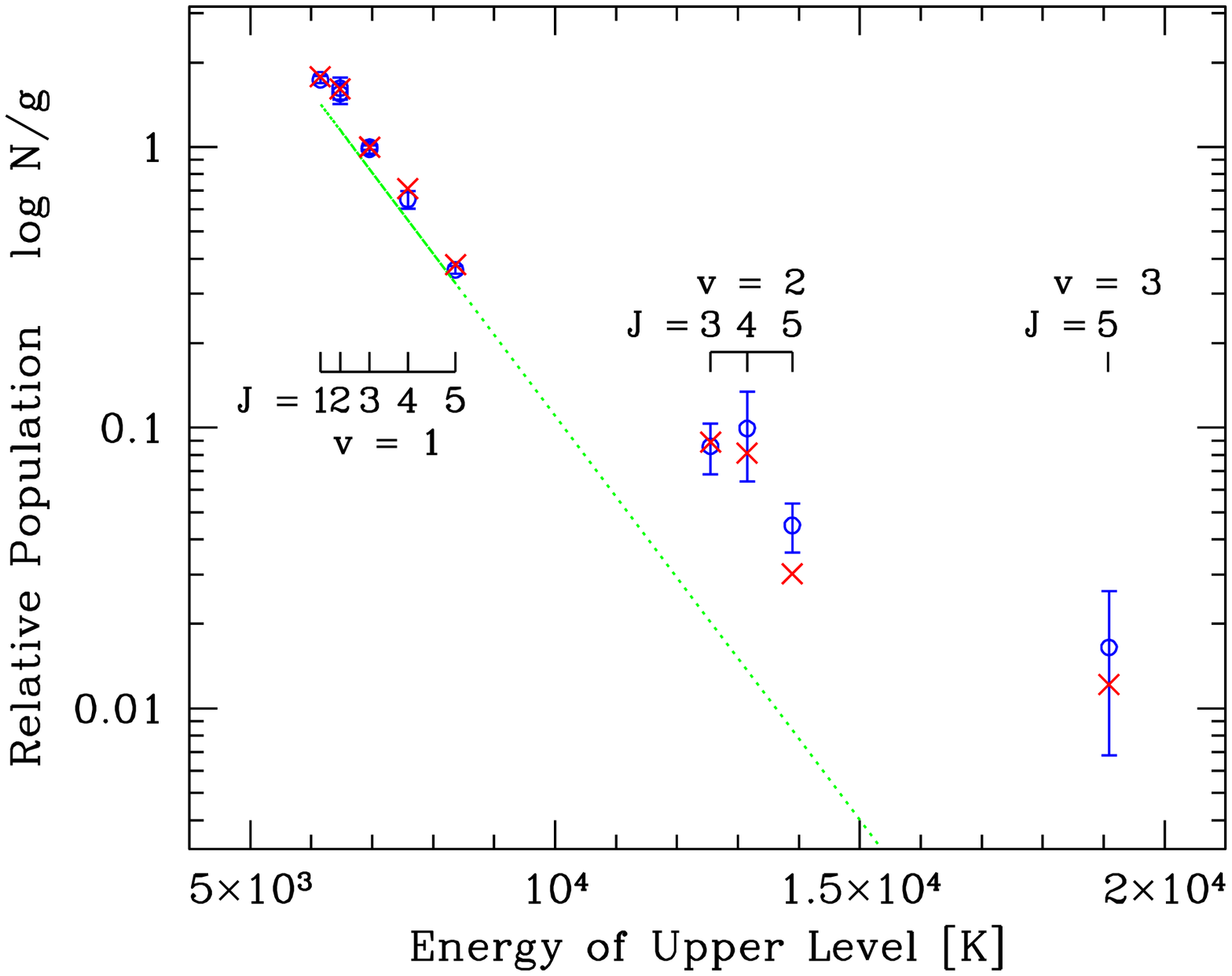,width=8cm}\psfig{file=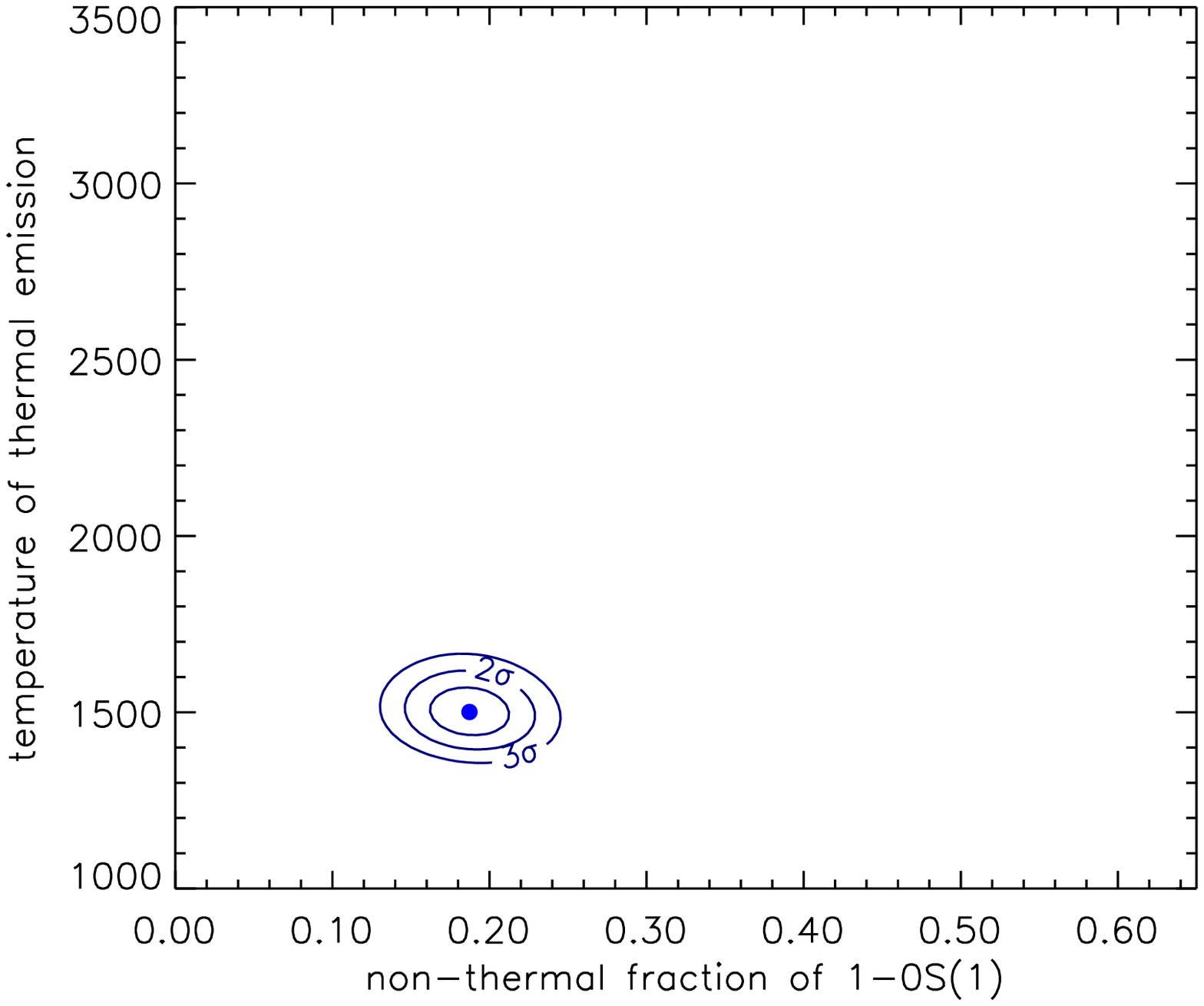,width=8cm}}
\caption{Left: Population diagram for the hot H$_2$ molecules in the NE
nucleus, derived for 11 line transitions.
The crosses denote the combined fit to the data of a thermal component
(dotted line, with
derived temperature $T = 1500$\,K) and a non-thermal component (Black \&
van Dishoeck (1987) UV fluorescent model 14).
The reduced chi-square $\chi^2_\nu = 0.89$, close to the expectation
value $\langle$0.92$\rangle$ for 8 degrees of freedom.
Right: Confidence regions for the thermal excitation
temperature, $T$, and the non-thermal fraction of 1-0\,S(1), $f_{UV}$.
The 1$\sigma$ uncertainties give $T = 1500\pm60$\,K and 
$f_{UV} = 0.19\pm0.02$ (ie fluorescent excitation accounts for 57\% of
the {\it total} H$_2$ line emission).\label{fig:h2pop_ne}}
\end{figure}

\begin{figure}
\centerline{\psfig{file=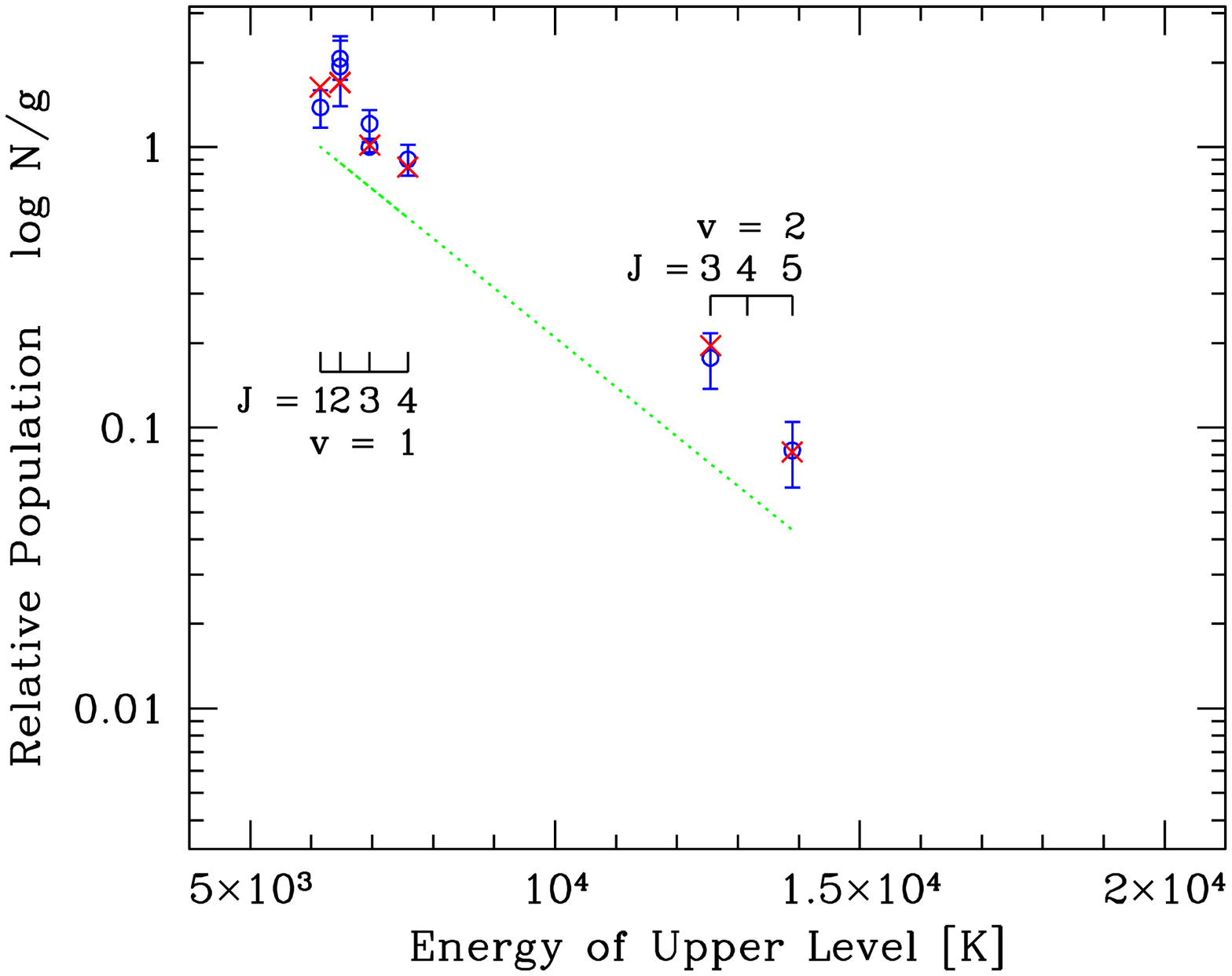,width=8cm}\psfig{file=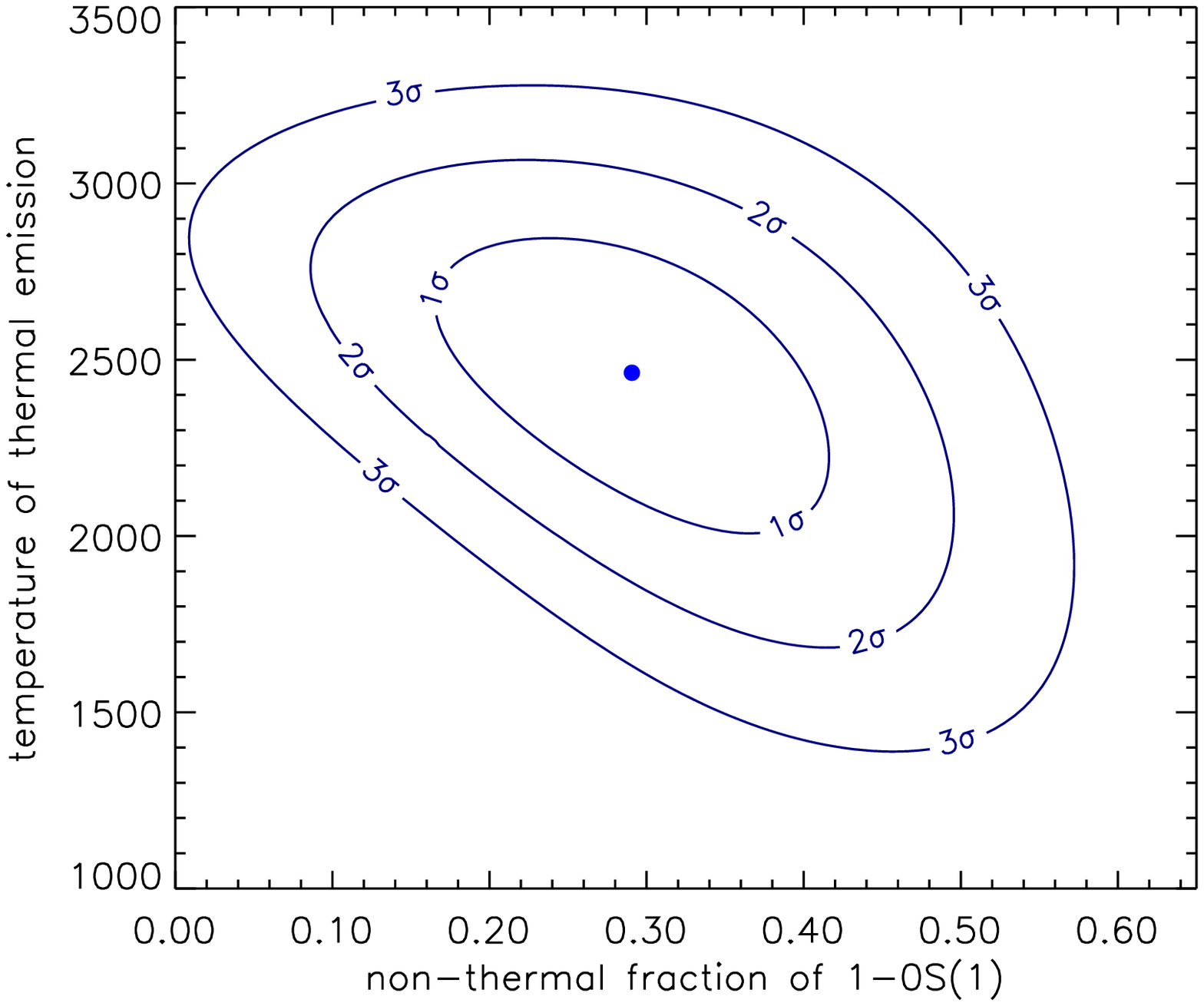,width=8cm}}
\caption{Left: Population diagram for the hot H$_2$ molecules in the SW
nucleus, derived for 8 line transitions.
The crosses denote the combined fit to the data of a thermal component
(dotted line, with derived temperature $T=2460$\,K) and a non-thermal
component (Black \& van Dishoeck (1987) UV fluorescent model 14).
The reduced chi-square $\chi^2_\nu=1.08$, close to the expectation
value $\langle$0.87$\rangle$ for 5 degrees of freedom.
Right: Confidence regions for the thermal excitation
temperature, $T$, and the non-thermal fraction of 1-0\,S(1), $f_{UV}$.
The 1$\sigma$ uncertainties give $T = 2460\pm410$\,K and 
$f_{UV} = 0.29\pm0.12$ (ie. fluorescent excitation accounts for 68\% of
the {\it total} H$_2$ line emission).\label{fig:h2pop_sw}}
\end{figure}

The H$_2$ line ratios relative to 1-0\,S(1) in each nucleus are given
in Table~\ref{tab:lines}, and level population diagrams are shown in
Figs.~\ref{fig:h2pop_ne} and~\ref{fig:h2pop_sw}, which have been derived
assuming the local thermal equilibrium ortho-para ratio of 3.
This should be valid for thermally excited H$_2$,
but it is generally found that for H$_2$ excited by UV fluorescence the
ortho-para ratio is $\sim$2.
Sternberg \& Neufeld (1999)\nocite{ste99} argue that this is because
the ortho absorption lines become optically thick sooner than the para
absorption lines,
so UV pumping of the ortho states is less efficient and
the ortho-para ratio appears to be reduced to a limiting case of
$\sim\sqrt{3}=1.7$.
That is, the UV excited states (those observed) have an apparent
ortho-para ratio of 1.7, while the total H$_2$ ratio remains at 3.
This effect only occurs if the absorption lines become
optically thick to UV pumping,
and is included in the line strength calculations of Black \& van~Dishoeck
(1987)\nocite{bla87} from whom we take our models of UV fluorescence.
However, for the purposes of fitting fluorescent models to the
population diagram, the ortho-para ratio used does not actually matter
as long as the same ratio is used 
for both the observed and modelled spectra when converting line
strengths to level populations.
For simplicity, therefore, since we do not know {\it a priori} the
relative thermal/non-thermal contributions (which is what determines the
observed ortho-para ratio), we have adopted the same ratio
as in the LTE case.

For one model, we have fitted a combination of thermal and non-thermal
components, with 
the 3 free parameters being the relative non-thermal contribution
$f_{\rm UV}$ to the 1-0\,S(1) line, the temperature $T$ of the thermal
component, and the absolute scaling.
For the non-thermal component we used fluorescent model 14 from Black
\& van~Dishoeck (1987), although this could equally well be formation
pumping which results in a similar spectrum.
For the thermal component we used the Boltzmann distribution at
temperature $T$ given by
\[
\frac{N_{\rm u}}{g} \ \propto \ e^{-E_{\rm u}/kT}
\ \ \ \ {\rm with} \ \ \ \ 
N_{\rm u} \ \propto \ \frac{F \lambda}{A_{\rm ul}}
\]
where $N_{\rm u}$ is the column density of H$_2$ in the upper level of
excitation energy $E_{\rm u}$, of an 
observed line which has wavelength $\lambda$, flux $F$, and transition
probability $A_{\rm ul}$ (quantities
are listed in Table~\ref{tab:lines});
$g$ is the degeneracy given by the product of the rotational and spin
degeneracies $g_{\rm J} \times g_{\rm S}$ where $g_{\rm J} = 2J+1$ and
$g_{\rm S} = 3$ for odd J and 1 for even J.
In the optically thin regime, these equations can be used to estimate
the mass of hot gas $M_{\rm H_2}$ from the line luminosity $L$ since
\[
L \ = \ 
f_{\rm u} \ A_{\rm ul} \ 
\frac{hc}{\lambda} \ 
\frac{M_{\rm H_2}}{m_{\rm H_2}}
\]
where $f_{\rm u}$ is the fraction of hot H$_2$ in the upper level of
the line 
% (cf Sugai et al. (1994)\nocite{sug94} who equate the line
% surface brightness with H$_2$ column density)
, and $m_{\rm H_2}$ is the mass of a hydrogen molecule.
Using the normalisation from Scoville et al. (1982)\nocite{sco82}
for gas thermalised at 2000\,K that for the 
$v=1$, $J=3$ level $f_{\rm u} = 0.0122$, the hot gas mass can be
estimated from the 1-0\,S(1) line luminosity as
\[
L_{\rm 1-0\,S(1)} (L_\odot) \ = \ 620 \ M_{\rm H_2} (M_\odot)
\]
In the population diagram, parameters were derived by minimising
chi-square; confidence regions were determined by systematically
varying each of $T$ and $f_{\rm UV}$ (which are not necessarily
independent) while simultaneously optimising the absolute scaling, 
until chi-square had increased by one standard deviation.
For the NE nucleus $T = 1500\pm60$\,K and $f_{\rm UV} = 0.19\pm0.02$; 
for the SW nucleus $T = 2460\pm410$\,K and $f_{\rm UV} = 0.29\pm0.12$.

For a second model we attempted to fit two thermal components, since
it is not implicitly ruled out by the data.
For the NE nucleus, the reduced chi-square $\chi^2_\nu=1.96$, 
rather larger than the expectation value of $\langle$0.91$\rangle$ for
7 degrees of freedom.
As well as this statistical arguement, there is a physical reason against
the model: the temperatures derived are 1200\,K and 5260\,K.
While the former only varies by $\pm$100\,K, the 1\,$\sigma$ limits on
the latter encompass the range 4000--7600\,K.
Since H$_2$ rapidly dissociates at temperatures above 4000\,K, it is
not possible to get a strong thermal spectrum at the
temperatures indicated here.
Thus we can rule out this model for the NE nucleus.
For the SW nucleus, the two-component thermal model effectively
reduces to a single thermal component which is itself simply a subset
of the previous thermal plus non-thermal model.
The value $\chi^2_\nu=1.95$ for a single thermal component supports the
result of the thermal plus non-thermal model above that a pure thermal
model is unlikely.

\subsubsection{UV Fluorescence}

If the non-thermal H$_2$ component arises in photo-dominated regions due to
UV fluorescence at the edge of H{\sc ii} regions, the models of Puxley
et al (1990)\nocite{pux90} can be used to constrain the compactness of the star
forming regions.
The diagnostic needed is the fluorescent 1-0\,S(1)/Br$\gamma$ ratio,
which we estimate to be $0.35\pm0.04$ (NE) and $0.16\pm0.07$ (SW);
the latter ratio may be intrinsically higher if any of the Br$\gamma$
is related to the AGN.

The ratios derived are rather higher than expected, and the models suggest
that they are typical of star formation which is either fairly diffuse
or relatively evolved (so the most massive ionising stars are no longer
present), both of which seem rather unlikely.
The former option contradicts observations that intense star formation
appears to occur mostly in compact clusters 
2--3\,pc across with masses up to 10$^6$\msun\ (eg Meurer et
al. 1995)\nocite{meu95} which would have ionising fluxes on the order of
10$^{52}$--10$^{53}$\,sec$^{-1}$ and hence, according to the model,
rather small ratios.
Indeed, Davies et al. (1998)\nocite{dav98b} found that the
1-0\,S(1)/Br$\gamma$ ratios of such clusters, averaged over the central
few tens of parsecs in the nuclei of blue compact dwarf galaxies, were
$\lesssim$0.1.
The latter option can be refuted by considering the Br$\gamma$
equivalent width, $W_{\rm Br\gamma}$, which implies a maximum age of
50\,Myr for the clusters dominating the 2\,$\mu$m continuum.
Since this time span is only greater than the main sequence lifetime of
stars more massive than 
$\sim$15\msun, many highly ionising stars will still be present,
implying a low UV flourescent 1-0\,S(1)/Br$\gamma$ ratio.
The high He{\sc i}/Br$\gamma$ ratios of $0.62\pm0.04$ and $0.47\pm0.03$
in the NE and SW nuclei respectively, also point towards very hot
stars.
Although not a unique indicator of stellar temperature (due to resonant
and, at high densities, collisional effects, Shields 1993\nocite{shi93}), such high
values suggest $T_{eff} = 30$--40$\times10^3$\,K equivalent to stars of
mass 15--30\msun.
As discussed below, some other process as well as UV fluorescence
is probably needed to account for much of the non-thermal H$_2$ emission.

At high densities, 
$\gtrsim10^5$\pcmcu, collisional effects will tend to thermalise the
H$_2$ molecules, altering the emission spectrum.
This means that in principle UV fluorescence can also give rise to
a spectrum that is apparently thermal, an effect used to explain the
anti-correlation between the ratio of 2-1\,S(1)/1-0\,S(1) and the
intensity of 1-0\,S(1) (Usuda et al. 1996)\nocite{usu96}.
Here, since it appears already that fluorescence can account for only
part of the weak non-thermal 1-0\,S(1) emission, it is unlikely to
contribute more than a few percent to the much stronger thermal part.

\subsubsection{X-ray Excitation}

In order to determine whether X-ray irradiation can be a potential
source of H$_2$ line excitation we need to estimate the 1--100\,keV
X-ray flux.
The 0.1--2.0\,keV X-ray luminosity has been determined by
Wang et al. (1997)\nocite{wan97} who analysed the X-ray emission from
Mkn\,266, making use of both the High Resolution Imager (HRI) and
Position Sensitive Proportional Counter (PSPC) instruments on 
{\em ROSAT} to identify a soft diffuse halo as well as 3 harder compact
sources. 
They found that one associated with the SW nucleus has a luminosity of 
1.3--$7.5\times 10^{41}$\,ergs\,s$^{-1}$ 
and the NE nucleus has 0.9--$5.1\times 10^{41}$\,ergs\,s$^{-1}$, 
depending on the absorbing hydrogen column density 
($N_{\rm H} =0.016$ \& $0.3 \times 10^{22}$\pcmsq).
However, they found that only thermal plasma models fit the data
satisfactorily, an unusual result given that the SW nucleus hosts an
AGN for which the canonical hard X-ray
spectrum is a power-law $F_\nu \propto \nu^\alpha$ with
index $\alpha = -0.7$.
That there is no way to be sure which sources seen with
the HRI correspond to which parts of the spectrum observed with the
PSPC adds confusion.
Since the SW nucleus contributes $\lesssim 20$\% of the total counts,
these may not have much impact on the 0.1--2\,keV spectrum, and
it is possible that the spectrum at energies lower than 2\,keV is
dominated by star-formation, while a Seyfert-like power-law would only
emerge at higher energies.
This is seen in some Seyfert\,2s, such as Mkn\,3 (Serlemitsos, Ptak \&
Yaqoob 1996)\nocite{ser96}.

This could well the case for the SW nucleus, where we make the reasonable
assumption that the AGN dominates the X-ray luminosity, so we can put
very comfortable upper limits on the extrapolated
1--100\,keV luminosity.
By normalising the 0.1--2\,keV region of a power-law model with the
same $N_{\rm H}$ to the entire HRI count rate of the source, we
can derive this upper limit as
$L \sim 12 \times 10^{41}$\,erg\,s$^{-1}$.
The specific case of X-ray illumination of dense tori around AGN was
modelled by Krolik \& Lepp (1989)\nocite{kro89}, who found that 
$L_{\rm 1-0\,S(1)} \sim 10^{-4} L_{\rm X-ray}$ for Seyfert 1s and 
$L_{\rm 1-0\,S(1)} \sim 10^{-5} L_{\rm X-ray}$ for Seyfert 2s.
The 1-0\,S(1) luminosity in a 5$''$ aperture (similar to that used to
estimate the X-ray flux from this source) around the SW nucleus in
our line image is around 2 orders of magnitude larger than predicted
for X-ray excitation alone.
An additional argument against significant X-ray excitation is the
offset of $\sim$500\,pc between the peak H$_2$ and AGN positions.

If the NE nucleus also hosted an AGN, similar arguments would apply.
If, on the other hand, we assume the X-rays are associated with
star-formation (eg supernovae or binaries), we should use the
luminosities directly from Wang et al. (1997)\nocite{wan97}.
The simplest models of H$_2$ for this case are those of Lepp \&
McCray 1983\nocite{lep83}.
They predict that the total 1-0\,S(1) luminosity is 0.25\% of the
X-ray luminosity, ie we would expect 0.6--$3\times10^5$\,L$_\odot$ of
1-0\,S(1) emission, whereas
the total luminosity observed in 5$''$ is $12.5\times10^5$\,L$_\odot$.
Thus X-ray excitation is expected to contribute 5--25\% of the
1-0\,S(1).

\subsubsection{Shock Excitation}

Shocks can be very effective at exciting H$_2$ molecules and here we
consider models of fast \& slow J \& C shocks from Burton et
al (1990)\nocite{bur90}.
These were originally tailored to the supernova remnant IC\,433, but the
general results have a much wider application.

\begin{description}

\item[fast J shock] (100--300\,km\,s$^{-1}$) would dissociate the molecules,
which then reform downstream on grain surfaces via formation pumping
(Mouri \& Taniguchi 1995)\nocite{mou95}.
The observed spectrum would be similar to that of UV fluorescence as the
molecules are created in energetic states which then decay radiatively
(see Table~1 in that paper).
These authors give the emissivity (photons emitted per molecule formed)
of 1-0\,S(1) as $\sim$0.02, requiring the shock(s) to pass through
1500\,M$_\odot$\,yr$^{-1}$ near each nucleus if 50--70\% of the
fluorescent emission is due to fast J shocks.
Additional tests are needed before we can confirm or rule out this process.
That an equal flux of Br$\gamma$ would be produced does not contradict
the observations, as long as some other process can make up the bulk of
the Br$\gamma$ without producing significant H$_2$ emission.
Observations of the J and H band [Fe\,{\sc ii}] lines would be useful, as
fast J shocks would destroy grains, enhancing the gas-phase Fe$^+$
abundance so that the line strength becomes comparable to 1-0\,S(1)
(Hollenbach \& McKee 1989)\nocite{hol89a}.

\item[slow J shock]
could be possible and would not dissociate the molecules.
Shock velocities in the range $v_{\rm s} = 5$--15\,km\,s$^{-1}$ 
result in strong H$_2$ lines and a post-shock temperature $T =
(3900$\,K)($v_{\rm s}$/10\,km\,s$^{-1}$) (Shull \& Draine 1987)\nocite{shu87}.
However, in normal ionisation fractions, magnetic field strengths, and
gas densities such shocks are expected to revert to C type
(Hollenbach, Chernoff \& McKee 1989)\nocite{hol89b}.

\item[fast C shock]
with $v_{\rm s} \sim 40$\,km\,s$^{-1}$ heats the gas to $\ge 2000$\,K
(dependent on $v_{\rm s}$) and produces strong H$_2$ line emission, which
could easily account for the bulk of the thermal excitation.
We estimate a mass of 1600 and 1000\msun\ of hot gas are needed near the
NE and SW nuclei.
The flow time through the hot shock structure of $\sim100$\,yr 
(Draine et al. 1983)\nocite{dra83}, then suggests that  
10 and 16\msun\,yr$^{-1}$ of gas must be shocked to sustain the
line's thermal luminosity.

\item[slow C shock]
is unlikely as the peak temperature is rather low, $\le 300$\,K,
resulting in very weak H$_2$ emission.
This would contradict the strong observed emission and derived
temperatures.

\end{description}

We have considered the various mechanisms which might give rise to the
H$_2$ emission in the vicinity of each nucleus ($R \lesssim 1$\,kpc), 
and find the simplest solution is that outlined below.
The non-thermal emission is a combination of UV fluorescence
(1-0\,S(1)/Br$\gamma \sim 0.1$) and fast 100--300\,km\,s$^{-1}$ dissociative J
shocks (1-0\,S(1)/Br$\gamma \sim 1$), the latter probably
accounting for 50--70\% with the ratios adopted here and heating
$\sim$1500\msun\,yr$^{-1}$ gas near each nucleus.
The thermal emission is due in part to X-ray heating but mostly fast
30--50\,km\,s$^{-1}$ non-dissociative C
shocks, heating only $\sim$10--16\msun\,yr$^{-1}$ of dense clouds.
In terms of the 1-0\,S(1) line, the contributions are approximately:
for the NE nucleus 5--25\% X-ray heating, $<$5\% UV fluorescence,
15--20\% fast J shocks, 55--75\% fast C shocks;
for the SW nucleus $<$20\% UV fluorescence, 10--30\% fast J shocks,
70\% fast C shocks.

\subsection{Spatial Analysis}

In order to reach sufficient signal-to-noise for the above analysis, we
examined only the integrated flux around the nuclei and ignored the
morphology.
A spatial anaylsis of the brighter features along the slit can also
be fruitful, as is apparent from Fig~\ref{fig:profile.s1brg}.
The upper row shows the continuum, 1-0\,S(1) line and Br$\gamma$ line
fluxes along the slit from the SW nucleus (left) to the NE nucleus
(right).
That the (relative) line fluxes do not exactly tally with those given in
Table~\ref{tab:basic} is due to different apertures: the data in
the table were extracted from a 5\arcsec\ circular aperture, those in
the figure from 0.61\arcsec$\times$1.22\arcsec\ boxes.
Also, no structures less than 1--1.5\arcsec\ will be visible, so the
details seen in the images in Fig~\ref{fig:morph} are not apparent.

The lower row of the figure is more interesting: it shows that the
ratios of 1-0\,S(1) to both Br$\gamma$ and the continuum {\it increase}
away from the nuclei.
Thus although the H$_2$ emission right in the nuclei may be
associated with Br$\gamma$ and the 2$\mu$m continuum (eg via star
formation processes), that more than $\sim$1\arcsec\ away is not.
Remarkably the ratios remain high even 2--3\arcsec\ away, a physcial
distance of 1--1.5\,kpc, although the surface brightness is very low.

\subsection{Extinction}

The extinctions derived from the H$\alpha$/H$\beta$
ratio in Veilleux et al. (1995)\nocite{vei95}, and also using the
H$\alpha$/Br$\gamma$ ratio, are very small (\av$\lesssim$3, Table~\ref{tab:basic}).
In the case of the NE nucleus, they are almost the same and suggest
a correct measure for the total extinction.
However, for the SW nucleus, the \av\ found using the Br$\gamma$
line is significantly larger, an indication that we are probing to greater
optical depths with the less-absorbed near-infrared line.
Additionally continuum fits to the spectra give A$_{\rm V}\sim4$,
remarkably consistent (expecially in the K-band where  A$_{\rm K} = 0.1
A_{\rm V}$) given the way they were determined, and that they probe the
stellar continuum rather than the excited gas.
Lastly, we measured the Pa$\alpha$/Br$\gamma$ ratio, which yielded
somewhat higher values of  A$_{\rm V}\sim7$ for the NE and 
A$_{\rm V}\sim13$ for the SW nuclei;
we must disregard these since the Pa$\alpha$ lines are seen at very low
($<$40\%) atmospheric transmission, and systematic errors of only
10--30\% would bring them in line with the other estimates.

\section{Discussion}
\label{sec:disc}

In this section we draw together the various lines of evidence that we
have examined, in an attempt to paint a consistent picture of the
events occuring in Mkn\,266.

\subsection{North East Nucleus}
\label{sec:disc_ne}

The NE nucleus is classified from its optical spectrum as a LINER by
Osterbrock \& Dahari (1983)\nocite{ost83} and as a Seyfert~2 by 
Veilleux et al. (1995)\nocite{vei95}.
Some confusion is unavoidable as 
explanations for LINER spectra are typically either weak AGN or shock
excitation, maybe due to a starburst (Filippenko 1996)\nocite{fil96}.
Both we (at 2$\mu$m) and MGAH (at 2--20\,cm) have
found that this nucleus is resolved on scales of several hundred
parsecs, making an AGN an unlikely proposition.
Additionally, Smith et al. (1998)\nocite{smi98a} found evidence for
simple structure at VLBI 18\,cm scales.
The flux they detected was 4.5\,mJy, only 20\% of that in the entire
nucleus, suggesting that even if there is an AGN component, it
certainly cannot dominate the power output.
Instead we assume that all the Br$\gamma$ is due to recent
star formation, and using models of Leitherer et
al. (1995)\nocite{lei95} we estimate the supernova rate 
($\nu_{\rm SN}$), K-band luminosity ($L_{\rm K}$), and bolometric
luminosity ($L_{\rm bol}$) with which it is likely to be associated.
We employ models with solar metallicity and a Salpeter (slope 2.35)
initial mass function (IMF) in the range 1--100\msun.

If active star formation occured over a short period and has since
evolved passively, then it must be very young as
Br$\gamma$ flux and equivalent width fall off very quickly with time;
the directly measured Br$\gamma$ equivalent width ($W_{Br\gamma}$) of
10\,\AA\ (in a 2\arcsec\ aperture) provides an upper limit to the age.
This is an unrealistic number to use since at least some of the
continuum is from an old underlying population (as evidenced by the
extended continuum):
decomposition of the deconvolved image suggests that in the central
2\arcsec, 
roughly 70\% of the continuum is from the core and the rest extended.
Additionally in Section~\ref{sec:contfit} we showed that supergiants
make up $\sim$25\% of the continuum.
Taking these two cases as the extremes implies that the $W_{Br\gamma}$
associated with the most recent episode of star formation should lie in
the range 15--40\AA.
This constraint give a range of 4.5--6.5\,Myr, consistent
with the CO index $CO_{\rm ph} = 0.12$. 
Such a young age implies 3--8$\times10^7$\msun\ of
stars would have had to form effectiv
ely instantaneously
($\lesssim1$\,Myr) in order to
reproduce the observed Br$\gamma$ flux.
We would then predict the parameters  
$\nu_{\rm SN} = 0.03$--0.08\,yr$^{-1}$,
$L_{\rm K} = 2$--7$\times10^8$\lsun, and 
$L_{\rm bol} = 3$--6$\times10^{10}$\lsun.

Models for continuous star formation may also be appropriate for 
this object, and
a plausible scenario might invoke `punctuated star formation'
in which clusters of 10$^5$--10$^6$\msun\ formed at different times over
this period.
$W_{Br\gamma} = 15$--40\,\AA\ then implies an age of 60--500\,Myr (or
more), during
which entire period the stars have formed with an average rate of
1.7\msun\,yr$^{-1}$ (ie, a few clusters every Myr).
60\,Myr is the time at which the maximum supernova rate 
$\nu_{\rm SN} = 0.03$\,yr$^{-1}$ is reached and thereafter sustained.
Unfortunately the CO index does not constrain the age range, although
both $L_{\rm K}$ and $L_{\rm bol}$ should since they increase
slowly with time as more late-type giants are accumulated.
Yet even after 500\,Myr the model predicts only 
$L_{\rm K} = 5\times10^8$\lsun\ and 
$L_{\rm bol} = 4\times10^{10}$\lsun.

The parameters for these two models are summarised in
Table~\ref{tab:sf_model} as Models~1 and~2 respectively, and are
compared with observed quantities.
For these we used the K-band luminosity from this paper and, as above,
suppose that somewhere in the range 25--70\% of that in 2\arcsec\ comes
from the starburst.
The bolometric luminosity is the infrared (8--1000\,$\mu$m) luminosity taken
from Sanders et al. (1991)\nocite{san91}.
We assume that each nucleus contributes in rough proportion to its
Br$\gamma$ and X-ray fluxes, amounting to 30--50\% for the NE nucleus.
The model predictions for these two quantities are factors of 3--4
less than those observed.

The supernova rate $\nu_{\rm SN}$ has an even larger discrepancy.
To estimate it from the 20\,cm continuum we used the
relation derived by Condon \& Yin (1990)\nocite{con90b}, between
non-thermal radio continuum and supernova rate.
By considering independent estimates of these for the Galaxy they found:
\[
L_{\rm NT} {\rm [W\,Hz^{-1}]}
\ \sim \ 
1.3\times10^{23} \ (\nu{\rm [GHz]})^{-0.8} \ \nu_{\rm SN} {\rm [yr^{-1}]} 
\]
This is very similar to the relation derived for M\,82 by Huang et
al. (1994)\nocite{hua94}, who find a coefficient of
$1.1\pm0.5\times10^{23}$ for that galaxy by considering the cumulative
number of sources as a function of increasing diameter.
Additionally, by using the temporal behaviour of the bright SN1979c in
M\,100 as a template, Colina \& P\'erez-Olea (1992)\nocite{col92}
derived a coefficient of $0.9\times10^{23}$ for the same equation.
So this relation appears to be quite robust, whether applied to the
Galaxy or to the rather different environment of starbursts.
From the 20\,cm radio continuum (MGAH) which has a
spectral index of 0.6 and hence will be
dominated by the non-thermal component at this wavelength, we derive 
$\nu_{\rm SN} = 0.45$\,yr$^{-1}$.

These calculations provide a strong indication that the starburst scaling
obtained from the Br$\gamma$ flux is too small by an order of magnitude.
There are potentially several ways around this problem.
A number of authors have suggested that the IMF might
be truncated or have a steep slope, both of which
reduce the number of most massive stars.
These two cases were also modelled by Leitherer et
al. (1995)\nocite{lei95}, and  
$W_{Br\gamma} = 20$\,\AA\ gives maximum ages of 30\,Myr (Miller-Scalo
IMF with slope 3.3) and 11\,Myr (upper mass limit set to 30\msun).
Although the normalisation of the model to the observed ionising flux
then increases, the predicted supernova rate actually decreases.
For the truncated IMF it is because the age is so small that many of
the progenitor stars are too young to produce supernovae;
for the steep IMF it is mainly because there are fewer supernova
progenitors.

The required effect of reducing the observed Br$\gamma$ flux could also
be achieved if there was dust internal to the ionised nebulae, which
then competes for Lyman continuum (Lyc) photons.
Perhaps more likely is that if the molecular clouds have been shredded
by multiple supernovae, the nebulae may be density bounded so that
many of the Lyc photons escape.
With the measured A$_{\rm V} \sim 2$ more than 90\% of these photons
would be mopped up by dust in the ISM, contributing to the
far-infrared flux without increasing the observed UV.
In either case the hydrogen recombination line strengths would be
reduced by the same factor.
An intriguing side-effect is to alter the ratio of H$\alpha$ or
H$\beta$ to the high ionisation species such as [O{\sc iii}].
Since lines with higher ionisation potentials are produced deeper in
the nebula, their strengths will not be affected if the nebula is
density bounded (except in rather extreme cases).
Such an effect could account for the high ratio of [O{\sc
iii}]/H$\beta$ measured by Veilleux et al. (1995)\nocite{vei95} leading
to their classification as a Seyfert.
We can hypothesise a `intrinsic' Br$\gamma$ flux (ie the
one we ought to measure if all the Lyc photons ionised H{\sc i}),
which would be many times larger than that observed.
However, this does not solve our problem since $W_{Br\gamma}$ must also
increase by the same factor, forcing us again to adopt a very young
starburst age.

One remaining hypothesis involving star formation is that 
at some point 10--30\,Myr ago a catastrophic event may have
triggered a single highly intense burst of activity, forming
$\sim7\times10^8$\msun\ during this period.
The starburst would now have evolved enough that it contributes little
to the current Br$\gamma$ luminosity (1--2\%) while accounting for the
radio continuum, $L_{\rm K}$, and $L_{\rm bol}$.
This is shown as Model~3 in Table~\ref{tab:sf_model}.
The corollary is that this leaves almost no age constraints on the star
formation responsible for the Br$\gamma$.
However, the difficulty here is that the radio and 2\,$\mu$m continua trace
out different regions so it seems unlikely they could originate from the
same episode of star formation.

A different approach is to consider an alternative origin for the 
synchrotron radiation -- which must nevertheless arise from electrons
accelerated by 
shocks since we have shown that an AGN is not a proposition.
Perhaps the two most important pieces of evidence are:
(1) the 1-0\,S(1) and radio continuum appear to be related to each
other because of their shock origin, and
(2) although the radio continuum is resolved, both it and the
1-0\,S(1) are very compact (on scales of 0.4\arcsec)
-- and, crucially, {\it more so} than the 2\,$\mu$m continuum (0.7\arcsec).
Our spectra show that the 2\,$\mu$m continuum is dominated by late-type
stars, and presumably traces out the region of recent star formation. 
We argue that perhaps the radio and 1-0\,S(1)
are {\it not} directly associated with the star formation.
Although the spatial scales are too small to make a definitive
statement, the morphology does appear to be similar to that of
NGC\,6240 where the gas has settled between the two continuum
nuclei and is radiating strong thermal 1-0\,S(1) emission 
(van der Werf et al. 1993, Sugai et al. 1997b, Tacconi et al. 1999)
\nocite{wer93,sug97b,tac99}.
If NGC\,6240 were moved away so that its nuclei, instead of being
1.5\arcsec\ apart, were separated by only 0.5\arcsec, it would begin
to look like Mkn\,266~NE.
We speculate whether Mkn\,266~NE could itself consist of two or more
(unresolved) continuum nuclei with gas settling between them.
As the models we have considered imply, star formation has occured
recently in these nuclei, but the supernova rate is very low so
essentially all the radio continuum originates in the settling gas.
A serious difficulty with such an interpretation is that
Mkn\,266~NE has a 3.6\,cm intensity (scaled from the 2\,cm, MGAH) of
7.0\,mJy within 0.3\arcsec$\times$0.4\arcsec, while NGC\,6240 has
diffuse 3.6\,cm emission at the level of 0.2\,mJy in a
0.23\arcsec$\times$0.32\arcsec\ beam 
(Colbert et al. 1994)\nocite{col94} -- some 35 times fainter.
Since we cannot invoke an AGN, it is hard to conceive of a method to
produce so much more non-thermal radio emission without reverting to
fast shocks.
Bell (1978)\nocite{bel78b} showed that the radio emissivity of shocked
gas is strongly dependent on the shock velocity $v_{\rm s}$:
\[
\epsilon(\nu) \ \propto \ 
\left(\frac{n_{\rm e}}{\rm cm^{-3}}\right)
 \ 
\left(\frac{B}{\rm 10^{-4}\,G}\right)^{\alpha+1}
 \ 
\left(\frac{v_{\rm s}}{\rm 10^4\,km\,s^{-1}}\right)^{4\alpha}
 \ 
\left(1 + 
  \left[\frac{v_{\rm s}}{\rm 7000\,km\,s^{-1}}\right]^{-2}\right)^\alpha 
\ 
\left(\frac{\nu}{\rm GHz}\right)^{-\alpha}
\]
So for a typical spectral index of $\alpha = 0.8$ we find that 
$ \epsilon(\nu) \propto v_{\rm s}^{1.6}$ for
$v_{\rm s} \lesssim 2000$\,km\,s$^{-1}$;
while 
$ \epsilon(\nu) \propto v_{\rm s}^{3.2}$
for $v_{\rm s} \gtrsim 7000$\,km\,s$^{-1}$.
To produce such intense synchrotron emission requires shocks with 
speeds of several thousand \,km\,s$^{-1}$, 3--10 times faster than
the 300\,km\,s$^{-1}$ seen in the CO line in NGC\,6240 by Tacconi et al. (1999)\nocite{tac99}.
However, in NGC\,6240 higher velocities than in the CO line were
observed in  1-0\,S(1) which has a 
FWHM of 550\,km\,s$^{-1}$ and FWZI of 1600\,km\,s$^{-1}$ (van der Werf et al. 1993)\nocite{wer93}.
Also, in Section~\ref{sec:h2excite} we saw that the fraction of H$_2$
excited by dissociation due to fast shocks is much higher in Mkn\,266
than in NGC\,6240; 
and to account for this $\sim$100 times more gas in Mkn\,266
is shocked at high velocities than at the low velocities expected in
dense molecular clouds.
Bell's equation shows that, keeping all other parameters the same,
increasing the shock velocity from 300\,km\,s$^{-1}$ to 3000\,km\,s$^{-1}$ would increase
the synchrotron intensity by a factor of 30, similar to the observed
difference between NGC\,6240 and Mkn\,266.
It does not seem totally unreasonable then to extrapolate from
NGC\,6240 to a more extreme environment in Mkn\,266~NE:
if gas has settled into a disk structure on scales several times
smaller than in NGC\,6240, the velocities required to support it
centrifugally will be many times greater and this could give rise to
the fast shocks required to produce the intense synchrotron emission.
High resolution observations of the CO in this nucleus will be needed
to decide whether this speculation is indeed borne out.

One final point to consider is the existence of faint but very extended
wings out to at least 1--1.5\,kpc in the 1-0\,S(1), as indicated in
Fig~\ref{fig:profile.s1}.
The lower limits to the S(1)/Br$\gamma$ ratio and the high S(1)
equivalent width indicated in Fig~\ref{fig:profile.s1brg} imply an
origin not associated with star formation.
A plausible alternative is that the lines are excited by shocks which
are driven into clouds in the turbulent medium of the X-ray halo.

\subsection{South West Nucleus}

This nucleus has also been identified as a LINER (Veilleux et al. 1995)
and Seyfert~2 (Osterbrock \& Dahari 1983).
The classification as a LINER rested solely on the ratio 
[O{\sc iii}]/H$\beta = 1.4$ because the diagrams used to distinguish
LINERs, Seyferts, and starbursts all plotted this value, against the
ratios of [N{\sc ii}], [S{\sc ii}], and [O{\sc i}] to H$\alpha$.
Evidence for an AGN comes from MGAH who observed an
unresolved radio core in the SW component using a beamwidth of
$0.3$\arcsec$\times0.4$\arcsec, and our deconvolved continuum image.
An additional line of reasoning put forward by Wang et al. (1997) is
that photoionisation by the hard radiation field of an AGN is needed to
explain both the luminosity of the large-scale optical emission line
nebula as well as the line ratios.
As discussed in Section~\ref{sec:images}, based on the 2\,$\mu$m continuum
we would argue for a combination of Seyfert plus starburst.

The surrounding emission, extending linearly about 1\arcsec\ both north
and south (radio as well as infrared) would be due to star
formation.
The Br$\gamma$ follows the distribution to some extent, although the
resolution is rather poor;
for example, the extension to the east is apparent in both continuum
and line images and may be due to an extranuclear H{\sc ii} region. 
Unfortunately, a quantitative analysis is not possible with the current
spatial resolution, since we cannot properly separate the AGN and
starburst contributions.

More interesting is the H$_2$, which peaks to the north east between
the galaxies in a region without any associated Br$\gamma$ or
continuum, and 500\,pc from the nucleus.
This would seem to rule out any connection with either the AGN or star
formation.
In Section~\ref{sec:h2excite} we showed that 70\% of the 1-0\,S(1)
emission in this nucleus was thermally excited and deduced that it
arose mostly in $\sim40$\,km\,s$^{-1}$ C shocks.
Together these suggest that it may due to shocks resulting from the
interaction being driven into clouds.
Both we and Veilleux et al. (1995) found a radial velocity difference
between the nuclei of 50\,km\,s$^{-1}$, setting a lower limit on the speed of the
interaction.
It is likely that the actual speed is faster, but shock speeds similar
to that required can be produced when a high speed shock in a low
density medium encounters a high density medium, the process favoured
for the H$_2$ excitation in NGC\,6240 by 
van der Werf et al. (1993)\nocite{wer93} and 
Sugai et al. (1997b)\nocite{sug97b}.
However, it is not possible to comment here on whether it is due to the
global interaction, or if it results from the gas dynamics more closely
associated with the SW nucleus.

\subsection{The Central Radio Continuum}
\label{sec:centre_bit}

One of the puzzling observations of Mkn\,266 is the
non-thermal radio continuum midway between the nuclei,
stronger than the extended emission in the SW nucleus and accounting
for more than 1/4 of the total 20\,cm emission (MGAH).
What makes this so surprising is that there is no evidence for any
other enhanced emission {\it of any sort} at this position.

MGAH argued that it could result from shocks as clouds collide near the
interface of the gas disks.
Our most severe observational constraint on this model is the
non-detection of 1-0\,S(1) in the locale, 
at a 5$\sigma$ limit of $10^{-18}$\wsqm\ 
(in a 5\arcsec\ aperture, similar in size to the
3\arcsec$\times$7\arcsec\ radio emission).
The MGAH scenario is ideally suited to the interacting galaxy
model of Jog and Solomon (1992)\nocite{jog92}, which describes pairs of
gas rich spiral galaxies colliding at 300\,km\,s$^{-1}$.
This model has an important addition to previous models (eg Harwit et
al. 1987\nocite{har87}) which have invoked
collisions between GMCs to explain the observations.
The authors note that GMCs have a low volume filling factor $f_{\rm
GMC}=0.01$ and so are unlikely to collide;
it is the H{\sc i} clouds with $f_{\rm H\sc i}=0.1$ which collide.
The GMCs are assumed to have a core radius of 25\,pc with an average
density $n_{\rm H_2}=100$\pcmcu;
the H{\sc i} clouds are given a radius of 5\,pc and an average density
$n_{\rm H}=20$\pcmcu, but with the larger filling factor, the total mass
is the same as the H$_2$.

The fast shocks in the H{\sc i} clouds are slowed in the denser GMCs
to $\sim 50$\,km\,s$^{-1}$, which are still nevertheless highly supersonic;
the compression timescale of $\sim2\times10^4$\,yr then implies that
in total only about 10\% of the GMC gas is shocked.
The cooling timescale is on the order of 100\,yrs (Drain, Roberge \&
Dalgarno 1983\nocite{dra83}), so only 0.5\% of the shocked H$_2$ is
radiating at any time.

At a distance of 115\,Mpc, and taking the line-of-sight dimension as an
average of the others, the volume over which the radio continuum is
emitted is 24\,kpc$^3$.
The H$_2$ mass inside this would then be $10^9$\msun, of which
$5\times10^5$\msun\ would be hot -- 
roughly the same amount as in NGC\,6240 (Sugai et al. 1997b),
and about 500 times more than is implied by our 5$\sigma$ upper limit
to the 1-0\,S(1) line.
Even allowing for large uncertainties in the parameters and timescales,
invoking shock excitation in clouds induced by the interaction appears
unlikely.

We offer a slightly different interpretation.
Wang et al. (1997) have argued that X-ray and optical emission line
observations point towards a radial outflow of gas at several hundred
\,km\,s$^{-1}$ which has swept up much of the ISM, and 
they conclude that the NE starburst and SW AGN play comparable
roles in photoionising the nebula.
At distances of a few kpc, the outflow velocity is $\sim300$\,km\,s$^{-1}$.
Our hypothesis is that the radio continuum traces the interface between
the expanding shock fronts from the two nuclei.
When added to the relative velocity of the galaxies, 
the relative velocities of these shocks will be approaching 1000\,km\,s$^{-1}$.
Copious quantities of non-thermal radio emission will be produced even
if the density of 
the shocked material is fairly low, but there will be no 1-0\,S(1) and
little hydrogen recombination flux as the gas is heated to
$10^6$--$10^7$\,K.
A test of this would be to observe the region with high spatial
resolution in the 0.1--2\,keV range, to see if there is a local
increase in the temperature of the X-ray emitting gas over the ambient
temperature in the bubble of 10$^6$\,K.

\section{H$_2$ Excitation: The Wider Perspective}
\label{sec:wide}

The classical case of H$_2$ emission occuring between the nuclei in a
merger is that of NGC\,6240 (Herbst et al. 1990\nocite{her90}, 
van der Werf et al 1993).
The H$_2$ level population is consistent with solely thermal
excitation and, combined with the absence of [Fe{\sc ii}] in the same
region, indicates that the {\it observed} emission arises from
non-dissociative shocks, occuring as a result of the merging process 
(Sugai et al. 1997b).
Recent CO radio observations (Tacconi et al. 1999)\nocite{tac99}
have confirmed the presence of 2--$4\times10^9$\msun\ of molecular gas
concentrated here.

Other radio CO observations of a number of galaxies at 1--3$''$
resolution by Gao et al. (1997) tend to
support this more generally, and
Scoville et al. (1997) argue that 2/3 of the
molecular gas in Arp\,220 relaxing into such a disk is consistent with
scenarios in which gas in merging systems settles into the centre
faster than the stellar nuclei.
They argue that efficient removal of angular momentum from the
observed dense centrifugally supported disk could play an important role
in the evolution of powerful starbursts/AGN.
They estimated a mass of $5.4\times10^{9}$\,M$_\odot$ in the disk with
a mean density of $2\times10^4$\,cm$^{-3}$.
Based on the model of Jog \& Solomon (1992) discussed in
Section~\ref{sec:centre_bit}, we estimated a ratio of cold to hot
(shocked) molecules of 2000.
This is not so different from the ratio of found in NGC\,6240:
Tacconi et al. (1999) estimated the cold H$_2$ mass to be 
2--$4\times10^9$\msun,
while Sugai et al. (1997b) found a cold H$_2$ mass of
$2\times10^5$\msun, yielding a ratio of $10^4$.
Even for this case the 1-0\,S(1) emission should be easily observable.
For example, for a merger at 200\,Mpc to produce a 1-0\,S(1) flux of
$10^{-18}$\,W\,m$^{-2}$, an H$_2$ disk mass of only
$\sim2\times10^7$\,M$_\odot$ is needed.
Given the examples of Arp\,220 and NGC\,6240, this seems a perfectly
feasible proposition for any 
of the luminous infrared galaxies (almost all at distances $<200$\,Mpc)
observed by  Sanders et al. (1991), in which the CO luminosity implies
total H$_2$ masses in the range $10^9$--$10^{10}$\msun.

We have now observed several other such systems similar to NGC\,6240 by
targeting close mergers, although with current observations it is not
possible to say whether the gas has formed a disk structure in these
galaxies.
These include Mkn\,551 (Sugai et al 1997a)\nocite{sug97c}, Mkn\,266 (here), and
Mkn\,496 (Sugai et al. 1999a)\nocite{sug99a}.
Also, the line images of NGC\,3256 (Kotilainen et al 1996) suggest that
there is an enhancement of 1-0\,S(1) between the two nuclei with
respect to both the Br$\gamma$ and [Fe{\sc ii}], which fall to a
minimum.
This phenomenon is not limited to the luminous infrared galaxies, and
has also been observed in some blue compact dwarf galaxies (II\,Zw\,40,
NGC\,5253, and He\,2-10, 
Davies et al. (1998)\nocite{dav98b}) where again the 1-0\,S(1) bears
little or no resemblence to the Br$\gamma$ or 2$\mu$m continuum.
However, the 1-0\,S(1) surface brightness is low in these interacting
systems, and it could only be detected because they are all nearby
($<$10\,Mpc). 
The more luminous galaxies with correspondingly larger gas masses do
form a better sample for further study, and
although such emission is not always observed (eg Arp\,299, Sugai et
al 1999b)\nocite{sug99b}, it may nevertheless be a relatively common
phenomenon and should be observable with the current generation of
infrared instrumentation.

\section{Summary \& Conclusions}
\label{sec:conc}

We have presented near infrared data on Mkn\,266 consisting of
2$\mu$m continuum, and Br$\gamma$ and 1-0\,S(1) emission line images,
as well as K-band spectra.
These are analysed in conjunction with data from the literature,
primarily the radio continuum (MGAH).

There are 3 main observations to note from the images.
\newline(1) 
In the NE (LINER) nucleus the 1-0\,S(1) is similar in size
($\lesssim$0.45\arcsec) to the radio continuum, but more compact
than the 2$\mu$m continuum (0.7\arcsec).
\newline(2) 
In the SE (AGN+starburst) nucleus, the 1-0\,S(1) is offset from the
continuum emission by 0.9\arcsec\ (500\,pc), and has no associated
Br$\gamma$ emission or radio continuum.
\newline(3) 
Midway between the two components where there is a region of strong
radio continuum, we find no enhanced near infrared line or
continuum emission at all. 

From our analysis of the spectra we note the following.
\newline(1) 
By fitting stellar templates to parts of the spectra away from emission
lines, we are able to remove all the absorption features, and hence
measure the fluxes of even weak emission lines.
Simultaneously we find that 20--25\% (NE) and 50\% (SW) of the
continuum is due to supergiants, the rest originating in late type
giant stars.
\newline(2)
Level population diagrams of the hot H$_2$ molecules indicate that
in the NE and SW nuclei respectively, 81$\pm$2\% and 71$\pm$12\% of the
1-0\,S(1) is thermally excited (to temperatures of 1500$\pm$60 and
2460$\pm$410\,K), mostly by $\sim$40\,km\,s$^{-1}$ C-shocks.
Additionally, much of the non-thermal emission may arise in faster
dissociative J-shocks.
Pure thermal models (with either one or two components of different
temperatures) are effectively ruled out.
\newline(3)
Although the 1-0\,S(1) surface brightness is very low away from the
nuclei, its ratio to Br$\gamma$ and the continuum remains high even
out to 2--3\arcsec.

The conclusions we draw are listed below.

{\bf NE nucleus}: 
the extended nature argues against an AGN, but
nor can any combination of star formation scenarios simultaneously
account for the observations.
Both the morphologies of the 1-0\,S(1) and 2$\mu$m continuum, and the
(predominantly) shock origin of the H$_2$ emission and radio
continuum, are reminiscent of NGC\,6240.
We speculate that this nucleus may resemble NGC\,6240 but on a smaller
physical scale, and with the higher velocities that would necessarily
occur in a smaller centrifugally supported disk.

{\bf SW nucleus}: 
the morphologies suggest an AGN with circumnuclear
star formation, but a quantitative analysis is not yet possible due to
the small angular scales involved.
Because the offset 1-0\,S(1) has no other associated emission
it cannot originate in star formation, and may instead be due to shocks
driven into clouds as a result of the interaction.

{\bf intermediate region}:
The lack of any enhanced 1-0\,S(1) emission suggests that the radio
continuum cannot trace out regions shocked by the interaction.
We propose that it indicates the interface between the expanding winds
from each nucleus, where combined shock velocities would be high enough
($\sim$1000\,km\,s$^{-1}$) that neither 1-0\,S(1) nor hydrogen recombination
lines would be observed.

\acknowledgments

We thank the staff at UKIRT and the JAC for their help during the
observing run.
Some of the data were obtained as part of the UKIRT Service Programme,
and we are grateful particularly to J. Davies who carried out the
observations.The United Kingdom Infrared Telescope is operated by the
Joint Astronomy Centre on behalf of the U.K. Particle Physics and
Astronomy Research Council.
RID acknowledges the support of the European Network on Laser Guide
Stars which operates under the auspices of the Training and Mobility of
Researchers programme.
The authors thank Matt Lehnert for stimulating and helpful discussions.

%References
\clearpage

\bibliography{/afs/mpa/home/davies/reference}
\bibliographystyle{/afs/mpa/data/irsi/v3/rid/styfiles/mnras/mnras}

%Tables
\clearpage

\begin{deluxetable}{rlcccc}
\tablecaption{Mkn\,266: Basic Quantities
\label{tab:basic}}
\tablehead{
\colhead{Quantity} &
\colhead{Unit} &
\multicolumn{2}{c}{North-East Nucleus} &
\multicolumn{2}{c}{South-West Nucleus}\\
\colhead{} &
\colhead{} &
\colhead{5\arcsec} &
\colhead{2\arcsec} &
\colhead{5\arcsec} &
\colhead{2\arcsec}
}
\startdata
2\,$\mu$m continuum\tablenotemark{a} & 
$\begin{array}{l}$[mJy]$\\$[mag]$\end{array}$ & 
$\begin{array}{c}  9.43\pm0.06 \\ 12.10\pm0.01 \end{array}$ & 
$\begin{array}{c}  4.09\pm0.03 \\ 13.00\pm0.01 \end{array}$ & 
$\begin{array}{c} 14.07\pm0.06 \\ 11.67\pm0.01 \end{array}$ & 
$\begin{array}{c}  5.76\pm0.03 \\ 12.63\pm0.01 \end{array}$ \\ 

Br$\gamma$ line\tablenotemark{a} & [$10^{-18}$\wsqm] & 
$3.4\pm0.4$ & $2.6\pm0.2$ & $5.7\pm0.4$ & $2.8\pm0.2$ \\

1-0\,S(1) line\tablenotemark{a} & [$10^{-18}$\wsqm] & 
$3.2\pm0.2$ & $2.2\pm0.1$ & $2.3\pm0.2$ & $0.7\pm0.1$ \\

1-0\,S(1)/Br$\gamma$\tablenotemark{a} & &
$0.94\pm0.13$ & $0.85\pm0.08$ & $0.40\pm0.05$ & $0.25\pm0.04$ \\

$W_{\rm Br\gamma}$\tablenotemark{a} & [\AA] &
$5.7\pm0.7$ & $9.9\pm0.8$ & $6.3\pm0.4$ & $7.6\pm0.5$ \\

\av \tablenotemark{b} & [mag] & 
    \multicolumn{2}{c}{2.3} & \multicolumn{2}{c}{0.9} \\

\av \tablenotemark{c} & [mag] & 
    \multicolumn{2}{c}{$2.7\pm0.2$} & \multicolumn{2}{c}{$2.1\pm0.2$} \\

\enddata
\tablenotetext{a}{Measured in aperture given;
errors are statistical, determined from noise in the images}
\tablenotetext{b}{from H$\alpha$/H$\beta$ ratio, 
Veilleux et al. (1995)\nocite{vei95}}
\tablenotetext{c}{from H$\alpha$/Br$\gamma$ ratio,  
where H$\alpha$ is from Veilleux et al. (1995) and 
Br$\gamma$ is measured in same region (3.7\arcsec\ length section of a
2\arcsec\ wide slit at position angle 36$^\circ$).
Errors are from Br$\gamma$ flux only.}
\end{deluxetable}

\clearpage

\begin{deluxetable}{rccccc}
\tablecaption{relative line strengths \& derived quantities\tablenotemark{a}
\label{tab:lines}}
\tablehead{
\colhead{Line} &
\colhead{Wavelength} &
\colhead{E$_{\rm u}/k$\tablenotemark{b}} &
\colhead{A$_{\rm ul}$\tablenotemark{b}} &
\multicolumn{2}{c}{Relative strength} \\
\colhead{} &
\colhead{$\mu$m} &
\colhead{K} &
\colhead{$10^{-7}$\,s$^{-1}$} &
\colhead{NE\tablenotemark{c}} &
\colhead{SW\tablenotemark{d}}
}
\startdata
1-0\,S(3)\tablenotemark{e} 
          & 1.9576 & \phn8365 & 4.21 & 0.76 &     ---  \\
1-0\,S(2) & 2.0338 & \phn7584 & 3.98 & 0.33 & \phn0.46 \\
2-1\,S(3) & 2.0735 &    13890 & 5.77 & 0.12 & \phn0.22 \\
1-0\,S(1) & 2.1218 & \phn6956 & 3.47 & 1.00 & \phn1.00 \\
2-1\,S(2) & 2.1542 &    13150 & 5.60 & 0.07 &     ---  \\
3-2\,S(3) & 2.2014 &    19086 & 5.63 & 0.04 &     ---  \\
1-0\,S(0) & 2.2235 & \phn6471 & 2.53 & 0.27 & \phn0.34 \\
2-1\,S(1) & 2.2477 &    12550 & 4.98 & 0.12 & \phn0.24 \\
1-0\,Q(1) & 2.4066 & \phn6149 & 4.29 & 0.81 & \phn0.65 \\
1-0\,Q(2) & 2.4134 & \phn6471 & 3.03 & 0.28 & \phn0.35 \\
1-0\,Q(3) & 2.4237 & \phn6956 & 2.78 & 0.69 & \phn0.85 \\
 \\
Pa$\alpha$\tablenotemark{f} 
          & 1.8756 &          &      & 5.32 &    15.00 \\
Br$\gamma$
          & 2.1661 &          &      & 0.54 & \phn1.82 \\
 \\
He\,{\sc i}
          & 2.0587 &          &      & 0.34 & \phn0.85 \\

\enddata
\tablenotetext{a}{Errors are estimated from the residual continuum near
each line. Due to poor atmospheric transmission and continuum
subtraction shortwards of 1.9$\mu$m, additional errors are present
in this region and we are unable to measure the Br$\delta$ (1.95$\mu$m)
or 1-0\,S(4) (1.89$\mu$m) lines.}
\tablenotetext{b}{For H$_2$ lines only; taken from the UKIRT
Astro-Utilities page: http://www.jach.hawaii.edu/JACpublic/UKIRT/astronomy/}
\tablenotetext{c}{Ratios are normalised to 1-0\,S(1);
1$\sigma$ errors are $\sim$0.02}
\tablenotetext{d}{Ratios are normalised to 1-0\,S(1);
1$\sigma$ errors are 0.05--0.07, 
 except 0.11 for the Q-branch}
\tablenotetext{e}{As there is no evidence for an AGN in the continuum,
we have assumed there is negligeable [Si\,{\sc vi}] emission at
1.96$\mu$m.}
\tablenotetext{f}{As explained in {\it a}, there may be a large
(10--30\%) systematic error on this line making extinction estimates
highly uncertain.}
\end{deluxetable}

\clearpage

\begin{deluxetable}{rlccccl}
\tablecaption{Mkn\,266~NE: Star formation models\tablenotemark{a}
\label{tab:sf_model}}
\tablehead{
\colhead{Quantity} &
\colhead{Unit} &
\colhead{Model 1\tablenotemark{b}} &
\colhead{Model 2\tablenotemark{b}} &
\colhead{Model 3\tablenotemark{c}} &
\colhead{Observed\tablenotemark{d}} &
\colhead{Note\tablenotemark{d}}
}
\startdata
Age           & $10^6$\,yr     & 4.5--6.5   & 60--500  & 20   & & \\
Stellar mass  & $10^7$\msun    & 3--8       & 10-85    & 70   & & \\
$\nu_{\rm SN}$& yr$^{-1}$      & 0.03--0.08 & 0.03     & 0.40 & 0.45   & 
from 20\,cm radio \\
$L_{\rm K}$   & $10^8$\lsun    & 2.1--6.9   & 2.7--6.0 & 20.6 & 6--18 & 
25--70\% of flux in 2\arcsec \\
$L_{\rm bol}$ & $10^{10}$\lsun & 3.3--5.7   & 2.9--3.8 & 12.0 & 9--15  & 
30--50\% of total $L_{\rm IR}$ \\
\enddata
\tablenotetext{a}{Using the models of Leitherer et al. (1995)}
\tablenotetext{b}{Instantaneous (1) and continuous (2) models
normalised to Br$\gamma$, assuming that 25\% (lower ages) 
and 70\% (higher ages) of the continuum within 2\arcsec\ is from the
starburst.}
\tablenotetext{c}{Instantaneous model (3) scaled to $\nu_{\rm SN}$, 
$L_{\rm K}$, and $L_{\rm bol}$.}
\tablenotetext{d}{The quantities measured, and where they came from}.
\end{deluxetable}

%Figures
\clearpage

\end{document}